\documentclass[aps,pre,superscriptaddress,amsmath,amssymb,floatfix,showkeys]{revtex4}
\usepackage{graphicx}
\usepackage{times}
\begin{document}
\def\D{\Delta}
\def\d{\delta}
\def\r{\rho}
\def\p{\pi}
\def\a{\alpha}
\def\g{\gamma}
\def\ra{\rightarrow}
\def\s{\sigma}
\def\b{\beta}
\def\e{\epsilon}
\def\G{\Gamma}
\def\om{\omega}
\def\l{\lambda}
\def\f{\phi}
\def\w{\psi}
\def\m{\mu}
\def\t{\tau}
\def\c{\chi}
 \title{A model for the self-organization of vesicular flux and protein distributions in the Golgi apparatus}

\author{Iaroslav Ispolatov}
\email{jaros007@gmail.com}
\affiliation{
Departamento de Fisica, Universidad de Santiago de Chile,
Casilla 302, Correo 2, Santiago, Chile}
\author{ Anne M\"usch}
\email{anne.muesch@einstein.yu.edu}
\affiliation{Department of Developmental and Molecular
Biology, Albert Einstein College of Medicine, The Bronx, NY, USA}

\begin{abstract}
The generation of two non-identical membrane compartments via
exchange of vesicles is considered to require two types of vesicles specified by
distinct cytosolic coats that selectively
recruit cargo and two membrane-bound SNARE
pairs that specify fusion and differ in their affinities for each type
of vesicles. 
The mammalian Golgi complex is composed of 6-8 non-identical cisternae that
undergo gradual maturation and replacement yet features only two SNARE
pairs. We present a model that explains how the distinct composition of Golgi cisternae
can be generated with two and even a single SNARE pair and one vesicle
coat. A decay of active SNARE concentration in aging
cisternae 
provides the seed for a 
{\it cis}$>${\it trans} SNARE gradient that generates the
predominantly retrograde
vesicle flux which further enhances the gradient. 
This flux in turn yields the observed inhomogeneous
steady-state distribution of Golgi enzymes, which compete with each
other and with the SNAREs for incorporation into transport
vesicles. We show analytically that the steady state SNARE concentration
decays exponentially with the cisterna number.  
Numerical solutions of rate equations reproduce the experimentally observed SNARE gradients, 
overlapping enzyme peaks in {\it cis}, medial and {\it trans}
and the reported change in vesicle nature across Golgi:
Vesicles originating from younger cisternae mostly contain  Golgi
enzymes and SNAREs enriched  in these cisternae and extensively recycle through the Endoplasmic Reticulum
(ER), while the other subpopulation of vesicles contains Golgi proteins prevalent in older
cisternae and hardly reaches the ER.
\end {abstract}


\maketitle

\section*{Author Summary}
We have developed a quantitative model to address a
fundamental question in cell biology: How does the Golgi apparatus, an
organelle composed of multiple cisternae that exchange
vesicles, steadily maintains its
inhomogeneous protein composition in the face of ongoing cisternal
aging and replacement, and cargo entry
and exit?
We do not assume any {\it a priori} polarity within the Golgi
apparatus or directionality of vesicular traffic. 
The Golgi cisternae inevitably lose active proteins that specify
vesicle fusion, the SNARE molecules, as they age, thus  breaking the
symmetry between compartments and establishing the "seed" for directional 
vesicular transport. 
This small decrease in SNARE concentration in older cisternae is then further self-enhanced
by the progressively more directional vesicular transport of SNAREs.
Competition of enzymes for incorporation into
predominantly retrograde-fusing vesicles 
in turn generates overlapping but distinct
stationary enzyme peaks. Applying these general mechanisms of fusion asymmetry
and competitive vesicle loading to the actual situation in the stacked
mammalian Golgi, we reproduced the experimentally
observed distributions of the two SNARE pairs that operate in the
Golgi, and enzyme peaks in {\it cis}, medial and {\it trans} cisternae. 
We believe that our study attempts the 
first self-consistent explanation for the establishment and
maintenance of
polarity in the Golgi stack.

\section{Introduction}
The Golgi apparatus is composed of multiple compartments, called
cisternae, typically 6-8 in mammalian cells. The individual cisternae
are enriched in glycosylation and other enzymes,
which form distinct but overlapping gradients with peaks in the {\it cis}, medial or {\it trans}
cisternae \cite{farquhar1981golgi}. 
\begin{figure}
\vspace{-1cm}
\begin{center}
\includegraphics[width=4in]{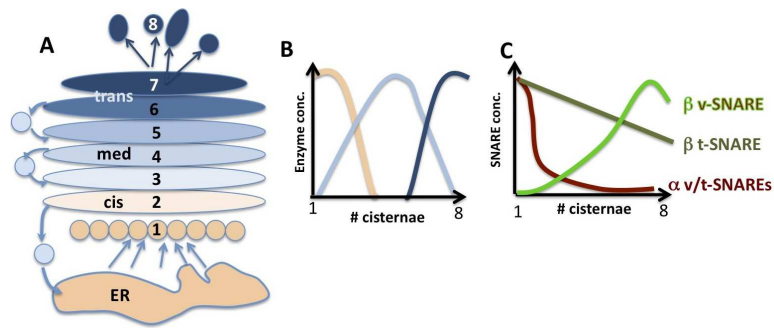}
\end{center}
\caption {\label{f1} 
{\bf Schematic representation of a stacked Golgi apparatus that undergoes cisternal maturation.}
A)   ER-derived vesicles (beige) fuse with each other to yield the
first, most {\it cis}, cisterna.  Individual
cisterna  mature from position 1 to position 8, where they
disintegrate into transport carriers destined for the plasma membrane
and endosomes. 
Vesicles originating from cisterna \#2
deliver {\it cis} Golgi proteins to cisterna \#1 while at the same
time cisterna \#2
receives Golgi resident proteins from cisterna \#3. 
B)   The cisternae are categorized as {\it cis}, medial and {\it trans} based on
the abundance of Golgi residence proteins, mostly glycosylating
enzymes, which exhibit distinct but overlapping peaks along the Golgi
stack according to their sequential role in the processing of exocytic
cargo. 
C)   Two SNARE pairs, which we term $\alpha$ SNARE (purple) and
$\beta$ SNARE (green)
are thought to mediate intra-Golgi transport of resident proteins. The
respective v and t-SNAREs of $\alpha$ SNARE both decay with a steep gradient
from {\it cis} to {\it trans}. $\beta$-t-SNAREs decay with a shallow gradient, while its
corresponding $\b$-v-SNARE concentration increases from cisternae 1 to
8. The graphs are schematic representations of data from
\cite{volchuk2004countercurrent}. 
}\end{figure}

 As anterograde cargo traverses the Golgi apparatus from {\it cis}
to {\it trans}, it becomes modified by Golgi enzymes in an assembly-line
fashion. Efficient and correct cargo processing depends on the distribution of glycosidases, glycosyltransferases
and other enzymes within the different Golgi sub-compartments in their
expected order of function \cite{roth2002protein}.
Several mechanisms for cargo movement through the Golgi
apparatus have been proposed. Of those, the cisternal maturation
hypothesis is best supported by all available experimental
data \cite{grasse1957ultrastructure}, \cite{glick2011models}. According to this concept,
cargo enters the Golgi by fusion of Endoplasmic Reticulum (ER)-derived vesicles with each other that form a
new cisterna at the {\it cis} face of the Golgi. The cargo exits the
Golgi in transport carriers that emerge from the {\it trans} most cisterna
when it disintegrates, thus maintaining the Golgi apparatus at a
steady state. Individual cisternae mature by shedding
their characteristic Golgi enzymes and at the same time acquiring
Golgi resident proteins from the more {\it trans} cisterna
\cite{glick1998curious}, \cite{pelham1998getting} (Fig.~1A).

It has been shown that Golgi resident proteins shuttle
between the cisternae in vesicles
\cite{martinez2001peri}, \cite{malsam2005golgin},
\cite{gilchrist2006quantitative}. But 
how  do individual cisternae acquire and maintain 
their specific and distinct enzyme compositions via vesicular
transport while the Golgi apparatus undergoes maturation?

Glick et al. provided one piece of explanation with a simple model
according to
which competition of Golgi proteins for incorporation into
retrograde-destined  vesicles accounts for their sorting within the Golgi
cisternae \cite{glick1997cisternal}. 
Proteins that are good competitors are efficiently
removed from the maturing cisternae and accumulate in the {\it cis} Golgi
while proteins that are poor competitors can only enter vesicles
after the good competitors have been depleted, and thereby end up in
more {\it trans} cisternae. While this model explains steady enzyme
segregation, it is based on an unexplained premise, namely, that the
Golgi-enzyme containing vesicles preferentially fuse with the younger
rather than the older cisternae.

Fusion of vesicles with acceptor membranes is specified by
\underline{S}oluble \underline{N}-ethyl-maleimide-sensitive factor  \underline{A}ttachment protein
\underline{Re}ceptors (SNAREs), integral membrane proteins that reside in the
vesicle and target membrane \cite{sollner1995snares},
\cite{jahn2006snares}. 
They function according to
a key-lock principle: Cognate SNAREs form a four-helical bundle, with
one chain contributed by a R-SNARE on one membrane and one heavy and
two light chains provided by corresponding Q-SNARE on the opposite
membrane to pull donor and acceptor membranes close enough to
fuse \cite{fasshauer1998conserved}. 
Theoretical work by Heinrich and Rapoport has shown that sets of
compatible SNAREs with preference for incorporation into a
specific type of coated vesicle can spontaneously generate and 
maintain non-identical compartments 
\cite{heinrich2005generation}  
when each compartment features a specific
pair of compatible SNAREs and corresponding vesicle type. The Golgi
however, maintains its 6-8 compartments with only 2 cognate SNARE
pairs and one type of vesicle (COPI) \cite{malsam2011organization}. 
How is this accomplished? A
higher concentration of SNARE complexes in younger compared to older
cisternae could readily explain the preference for retrograde fusion
of COPI vesicles, which in turn can yield the differential enzyme
peaks as described by Glick et al. \cite{glick1997cisternal}.
A {\it cis}-to-{\it trans} decrease is indeed
observed for Golgi Q-SNAREs \cite{malsam2011organization} 
(Fig.~1C). But how are these SNARE gradients
established in the first place?  To complicate matters, a R-SNARE
implicated in intra-Golgi traffic forms a counter-current gradient
with increasing levels from {\it cis}-to-{\it trans}, 
\cite{banfield1995snare} \cite{volchuk2004countercurrent},
(Fig.~1C). How is this
compatible with retrograde transport?

We present a model  of inter-cisternal
vesicular transport in which we do not
assume any  {\it a priori}
asymmetry within the Golgi apparatus. The transport is mediated by 2 cognate SNARE pairs, which compete
with each other and with other Golgi residents for incorporation into
a single vesicle type. 
The retrograde directionality
of vesicular flux is triggered by
the
temporal decrease of the concentration of cisternal SNAREs, which occurs via loss of
SNARE-containing vesicles,
including the recycling of COPI vesicles from the Golgi to the ER), 
decay, and inhibition of SNARE molecules.  As a result,  cisternal age becomes a distinguishing factor: {\it trans} cisternae are
older than {\it cis} cisternae and thus contain less SNAREs. 
A small distinction in SNARE concentrations provides the
seed for a {\it cis}$>${\it trans} gradient, which becomes self-enhanced by 
vesicular transport of the SNAREs.   
The steady SNARE gradient
controls a predominantly retrograde vesicular flux in which Golgi enzymes with stronger
affinities for the coated vesicles cycle predominantly between the {\it cis}
cisternae and the ER, while weaker-binding enzymes only enter
vesicles from later cisternae and exhibit less ER retrieval.  

\section{Results}

\subsection{General features of the model}
We assume that\\*
1. The Golgi consists of a stack of $n$ cisternae, which
move in an anterograde direction or ``mature'',
carrying  with them their SNAREs, enzymes (such as glycosyltransferases),  and  proteins that are
being processed.  The latter will not
be considered here.  Once every
$\tau$ time units, a new cisterna is added to the {\it cis} end of
the stack, while the most mature cisterna dissolves and disappears from the {\it
  trans} end of the stack.  The new cisterna is formed
by coalescence of ER-derived vesicles and  contains fixed
concentrations of SNAREs and enzymes.
\\*
2. Along with cisternal progression,  vesicles containing SNAREs  and
Golgi enzymes  continuously
bud from each cisterna.  We assume that the vesicles provide local
transport and can only fuse with the neighboring {\it cis} (less mature) and {\it trans}
(more mature) cisternae, and with the progenitor. 
Indeed, in the stacked mammalian Golgi, coil-coiled vesicular
tethering factors which span the distance between adjacent
cisterna  are thought to grab vesicles even prior to
their release from the donor cisterna and prevent them from reaching
more distant cisternae \cite{lupashin2005golgi}, \cite{guo2008coat}.
We will later relax this restriction and consider transport in a 
non-stacked Golgi, as it exists, for example, in the yeast {\it
  Saccharomyces cerevisiae}.
\\* 
3. SNAREs and Golgi enzymes are uploaded into a
 vesicle via competitive binding  to a fixed number of vesicular
 sites. We assume that the vesicular transport results primarily
  in the movement of cargo without any significant change in the
  volume and budding surface area of the cisternae.  This is supported
  by observations that the size of all cisternae is similar  \cite{ladinsky1999golgi} and our
  estimates that taking into account the vesicular transport of membrane
  itself would not significantly alter the results. \\*

Functioning of the model hinges on two general principles:
Establishment and maintenance of a directed retrograde vesicular flux and sorting
of the vesicular cargo via competition for binding sites.  

\subsection{Establishment of a {\it cis}$>${\it trans} SNARE gradient that mediates
 retrograde vesicular flow} 
To reveal the
universality of the proposed self-establishing mechanism of vesicular traffic
directionality we  first consider the simplest possible setup, a single cognate SNARE pair and vesicle type. 
We assume that the rate of vesicular fusion is proportional to the
product of the concentrations of the SNAREs present in vesicles and cisternae, respectively. The precise
nature of SNARE molecules  does not have to be specified here. We can
even consider the SNAREs as mere proxy for fusion-specifying factors.
The probability for a vesicle to fuse with a given cisterna depends
solely on the cisternal concentration of compatible SNAREs, and
cisternae with higher SNARE concentration have a higher probability to
absorb vesicles.   A retrograde vesicular flux 
thus requires a {\it cis}$>${\it trans} gradient in cisternal SNAREs. 

We propose that key to a robust {\it cis}$>${\it trans} SNARE gradient
is the observation that all systems, living and otherwise, function with
a loss. As Golgi cisternae mature they 
inevitably lose active SNARE molecules. Such a decay of active
SNAREs breaks the 
symmetry between the otherwise identical cisternae in a systematic way: 
The older {\it trans} cisternae contain less SNAREs than the younger {\it cis}
cisternae. The SNARE loss can occur
by escape of SNARE-carrying
vesicles that fuse with the ER thus recycling their
content. However,  some of the cisternal SNARE decay is likely due to
irreversible loss that requires some new SNARE synthesis to replenish
the system.

The ``seed'' SNARE gradient generated in this manner sets a preference for vesicles
to fuse with {\it cis} rather than {\it trans} cisternae, thus initiating the
directed vesicular transport. As SNAREs are transported retrograde,  
their {\it cis}$>${\it trans} gradient is further
enhanced. When the vesicular flux
becomes balanced by the 
anterograde transport  of SNAREs due to cisternal maturation, the system comes
to a steady state.
\begin{figure}
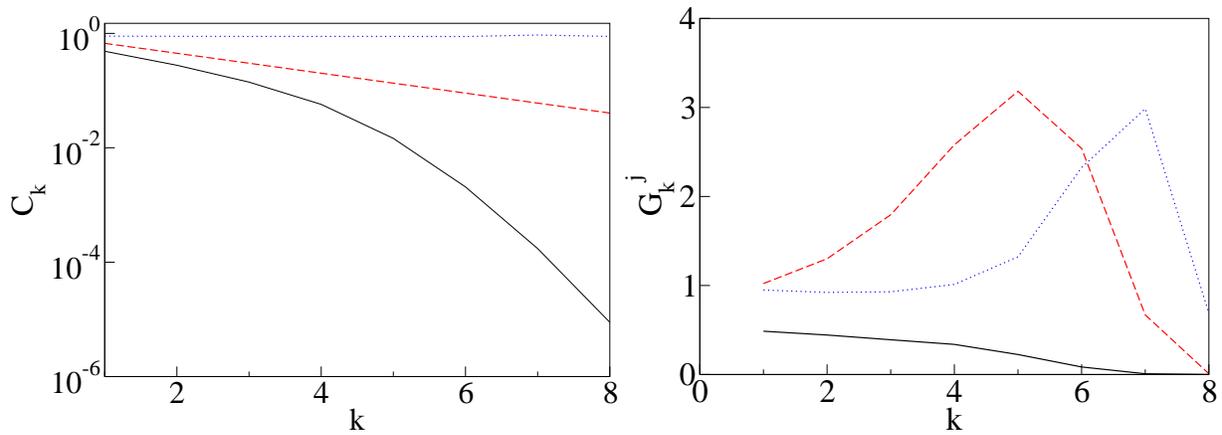

\begin{center}
\includegraphics[width=0.45 \textwidth]{f2a.eps}
\includegraphics[width=0.44 \textwidth]{f2b.eps}
\end{center}
\caption{\label{f2}{\bf Self-generated concentrations of SNAREs and enzymes.}
{\bf Left panel:} Panel A: Steady state concentration
 of cisternal SNARE $T_k$ vs the number of cisterna $k$ for: both the
 loss  and the vesicular transport mechanisms are enacted (solid
 line), only the loss mechanism operates (dashed line),  only
 vesicular transport occurs (dotted line). All concentrations are sampled immediately before the
cisternal shift event, when the number of each cisterna is incremented
by one. The definitions of parameters are given in Methods. 
Here and in all following plots it is assumed that  $T_0=1$
and $\tau=1$, i.e. all concentrations are expressed in the units of
initial concentrations and the time is expressed in units of the cisternal
maturation period. Solid line: $\eta \tau
=0.4$ and $\gamma \beta S B \tau=1.5$, dashed line:
$\eta \tau
=0.4$ and $\gamma \beta S B \tau =0$,  dotted line: 
$\eta \tau
=0$ and $\gamma  \beta S B \tau=0.5$, for all curves
$K=0.5$.
 {\bf Right panel:}
  Distribution of Golgi enzymes: {\it cis} (solid line), medial (dashed line) and {\it trans}
  (dotted line) established as a result of competition
   for incorporation into vesicles. 
  Vesicular flux is controlled by the gradient of cisternal SNAREs shown by the
  solid line in the left panel, vesicles from
  the first cisterna can exit the Golgi and fuse with the ER.  The
  parameters for the enzyme transport are
  $\gamma \beta S \tau =6$,  $K_1=0.4$, $K_2=0.6$, and
  $K_3=1.8$. 
}\end{figure}
Indeed, we show both numerically and analytically,
Figs~\ref{f2} and \ref{f5} and
Eqs.~(\ref{lambda}, \ref{form3}, \ref{small_loss})
that
the seed gradient, created by the temporal decay of
SNAREs, is self-enhancing.  Importantly, while
the vesicular transport significantly increases the seed gradient
produced by SNARE loss, without the loss the  vesicular transport by itself cannot
produce or maintain any gradient,
see Eq.~(\ref{small_loss}) and subsequent illustrations in Methods.
 This is in accordance with the results of \cite{heinrich2005generation}
that the single SNARE pair/single
coat minimal system cannot spontaneously break the initial symmetry of
compartments. The constant progression of cisternae is equally
important for maintaing the steady state SNARE gradient and
directional vesicular flux. Without the progression, the seed SNARE
gradient would have been equilibrated via vesicular transport. 

We note that at steady state the vesicular flux does not
depend on the concentration of SNAREs in the vesicles: Lower
concentrations of vesicular SNAREs are compensated by a higher steady
state number of vesicles.   Naturally, a vesicle should contain a minimum number of
vesicular SNARE molecules to ensure any  fusion at all.

The calculation of the steady state SNARE gradient and vesicular flow are presented
in the Methods section.

\subsection{ Establishment of Golgi enzyme peaks in {\it cis}, medial and {\it trans} cisternae via SNARE-mediated retrograde vesicular traffic}
 Next, we investigated how retrograde vesicular flow, created by the
 cisternal SNARE gradient, maintains the inhomogeneous steady state
 distribution of Golgi enzymes during cisternal maturation. To this
 end, we further developed the principle proposed by Glick et
 al. that attributes the different cisternal
 enzyme profiles to the competition of enzymes for the binding sites
 in vesicles \cite{glick1997cisternal}.
For simplicity, we assume three categories of Golgi enzymes with
 peaks in {\it cis}, medial and {\it trans} cisternae, and with
 strong, intermediate and weak affinities for vesicular
 binding sites, respectively (Schematically depicted in
 Fig.~1B).  Unlike in earlier models
 \cite{glick1997cisternal} and \cite{weiss2000protein},
 the fraction of binding sites occupied by each
type of enzyme is determined by  mass action equilibrium. 
Also, in contrast to  \cite{glick1997cisternal} and \cite{weiss2000protein}
where a number of {\it ad hoc} assumptions
about vesicular flow were used,  we ``couple'' the  enzyme-carrying capacity
to  the self-established vesicular flow described above. 
Hence, while each vesicle 
competitively uploads enzymes according to their dissociation
constants, its  fusion probability is determined by
the cisternal SNARE gradient shown by the black curve in the left
panel of Fig.~\ref{f2}. To study the competition mechanism in its
simplest form, we assume here that the SNARE
distribution is unperturbed by enzyme uploading.

We find that the distribution of enzymes radically depends on whether vesicles originating from the
first cisterna can exit the Golgi and fuse with its {\it cis} neighbor, the ER. If we permit ER recycling, 
the desired {\it cis}-medial-{\it trans} 3-mode
steady state localization can be reproduced (right panel of Fig.~\ref{f2}).
Enzymes that have the highest affinity for the vesicular coat concentrate
in the {\it cis} Golgi. A substantial fraction of these enzymes is loaded
from the first cisterna into ER-bound vesicles and leaves the Golgi. In
more central compartments, where {\it cis} enzymes are depleted, medial Golgi
enzymes outcompete the weaker-binding {\it trans} enzymes for space in the
vesicles. As a result, those enzymes advance with the maturing
cisterna until the mid-Golgi where their  concentration peaks, and
then become effectively loaded into retrograde vesicles. 
Finally, the weak-binding enzymes can only incorporate into
vesicles when all stronger-binding competitors are depleted. Their
concentration peaks in the penultimate cisterna. The ultimate
cisterna, equivalent to the disintegrating cisterna or {\it trans} Golgi
network (TGN), exhibits a somewhat lower enzyme concentration as it
does not receive any incoming retrograde vesicular traffic.

On the
other hand,  if none of the enzyme-carrying vesicles can escape to the ER, the cisternal distribution of all enzymes converges onto a
single peak form (see Fig.~\ref{f6}, Methods). 
In this case the overall steady state
abundance of enzymes increases with their affinity for vesicular
binding: The stronger-binding enzymes are more efficiently retrieved
to younger cisternae and thus better avoid being flushed out with the disintegrating
{\it trans} cisterna than the weaker-binding enzymes. As a
consequence of their higher concentration, the stronger-binding
enzymes do not get sufficiently removed from younger cisternae to
achieve their {\it cis}-Golgi peak and at the same time they do not give
their weaker competitors any chance to enter the retrograde transport
vesicles in the later cisternae. Hence, all enzymes peak at the {\it trans}
face of the Golgi. 
Thus, recycling of enzymes to the ER is necessary
for establishing the {\it cis}-medial-{\it trans} enzyme segregations. 
At the same time, we observe that the steady state SNARE distribution and resulting intra-Golgi vesicular
flux is  only weakly affected by the presence or absence of ER-recycling. This is because both scenarios feature an inherent loss mechanism, which breaks the intra Golgi conservation of SNAREs. 

We also observe that, as discussed in \cite{weiss2000protein},
the competition-based enzyme segregation is rather sensitive to the variation of
model parameters. Thus, it is possible that mechanisms have evolved to make
the cisternal enzyme distribution more robust. One such mechanism, the
change in enzyme affinity for vesicular binding sites with cisternal age,
has been studied in \cite{weiss2000protein} 
and could easily be incorporated into the a more detailed versions of our model.

The quantitative details of the calculation of the steady state enzyme
concentrations are presented
in Methods.

\subsection{The two Golgi SNARE pairs can function with a single vesicle type to establish their own gradients and the observed Golgi enzyme peaks in {\it cis}, medial and {\it trans}}
We now apply the general mechanisms of fusion asymmetry and competitive vesicle binding to explain the specific SNARE and enzyme distributions as they are actually observed in the mammalian Golgi. The important adjustment to our basic model is that the Golgi apparatus features not one, but two cognate SNARE pairs. 
The first
pair, which we label $\a$,
consists of the monomer SNARE rBet1 with its trimer SNARE partner
Membrin/ERS24/Syntaxin5. The second pair, labeled $\b$, consists of the monomer SNARE
GS15, compatible with the trimer SNARE complex of Gos28, Ykt6 and
Syntaxin5. There is solid experimental evidence for both pairs to
be incorporated in COPI vesicles \cite {volchuk2004countercurrent}, \cite{cosson2005dynamic}
and to participate in vesicular traffic of Golgi resident proteins 
 \cite{wooding1998dynamics}, \cite{xu2002gs15}, \cite{bruinsma2004retrograde}.

To reproduce three Golgi enzyme peaks in
concurrence with the experimentally observed distributions of the
$\alpha$ and $\b$ SNARE pairs we introduce an additional specification at
this point, namely that the monomeric SNAREs rBet1 and GS15 mediate
fusion only when present on the vesicle, and the trimeric-SNARE
complexes only when present in the cisternae. 
In the following paragraph we provide a justification for the
functional allocation of SNAREs as vesicular and cisternal. 

In the Golgi, only the $\a$ SNARE proteins
actually have  a {\it cis}$>${\it trans} distribution \cite
{volchuk2004countercurrent} such as shown in Fig.~2.
The $\b$ SNARE Gos28 also
decreases from {\it cis} to {\it trans}
\cite{subramaniam1995monoclonal};
however, its cognate monomeric SNARE
partner GS15 accumulates in the {\it trans}-most cisternae instead 
\cite {volchuk2004countercurrent},
and the GS15 yeast homologue Sft1p is also enriched in the late Golgi
\cite{banfield1995snare}, \cite{wooding1998dynamics},
(see Fig.~1C).
If GS15 and the Gos28-Ykt6-Synt5 complex could
both function as fusiogenic SNAREs in the cisternae, our model of
vesicular flux would imply that Golgi enzymes known to depend on this
SNARE pair for vesicular traffic undergo anterograde rather than
retrograde transport. The anterograde vesicular enzyme
transport does little to improve enzyme segregation as the cisternal
maturation already  moves enzymes in {\it trans} direction. More importantly,
the anterograde vesicular transport makes 
the enzyme recycling impossible.  Our allocation
agree with {\it in vivo} observations: monomeric SNAREs act
indeed most often as vesicle- or v-SNAREs and the trimeric SNAREs
generally function at the target membrane (and are therefore typically
referred to as t-SNAREs),  \cite{malsam2008membrane}.
But we also have a mechanistic explanation
for why trimeric Golgi SNAREs function in the cisternae rather than
the vesicles: When we consider the monomeric and trimeric SNAREs of a
cognate SNARE pair separately, the SNARE that is most abundant in the
vesicle determines which of the cisternal SNAREs the vesicle
engages with. If the monomeric SNARE is more abundant in a vesicle than the
trimeric SNARE, it will specify that the vesicle fuses with the
cisterna which has the highest amount of cognate trimeric SNAREs,
regardless of its monomeric SNARE concentration. Thus, when monomeric
and trimeric SNARE partners differ significantly in their affinity for
vesicles, the one with higher affinity becomes the v-SNARE, leaving
the other to function in the cisternae. This is the case especially
for the $\b$
SNARE pair as 
Syntaxin 5, the limiting partner in both $\a$ and $\b$ trimeric SNARE
complexes, is  at least 4 times less abundant than the $\b$ monomer GS15 in COPI
vesicles (See Fig.~7B in \cite{volchuk2004countercurrent}). 
Syntaxin-5's
apparent poor affinity for Golgi vesicles explains its observed
homogenous distribution in Golgi cisternae. However, the other constituents
in $\a$ and $\b$ trimers are more efficiently transported by vesicles,
thus maintaining the cisternal SNARE gradient.
It follows from these observations that
the two Syntaxin 5 containing trimeric Golgi SNAREs function as
t-SNAREs.
  
In addition to the two Golgi SNAREs, we consider a third v-SNARE,
which mediates the fusion of Golgi-derived vesicles with the ER. It is
ERS24, which thus has a dual function as  part of a
t-SNARE complex in intra Golgi transport and as v-SNARE in Golgi-to-ER
transport. The corresponding ER t-SNARE 
does not leave the ER and is therefore not considered here \cite{malsam2011organization}.

Apart from the SNARE specifications, we implemented a
similar set of minimal assumptions as for the single SNARE scenario: \\*
1. The rate of vesicular fusion with Golgi cisternae is determined
  by both  $\a$ and $\b$ SNARE pairs and is proportional to the
  sum of the products of the  cognate v-SNARE and t-SNARE
  concentrations. \\*
2. All SNAREs and Golgi enzymes
  compete for the same binding sites in the vesicles. This is in
  agreement with the findings for ER-derived COPII vesicles, the only
  instance where cargo-competition for coat binding has been
  elucidated to date \cite{miller2003multiple}. \\*  However, our model reproduces the enzyme segregation as well in the case  when the enzymes compete only with each other and not with SNAREs for vesicular binding sites, as shown in Fig.~\ref{f2}. 
 
3. Vesicles fuse locally, i.e. with the {\it cis} and {\it trans} neighbors of
  the progenitor cisterna and the progenitor cisterna itself. In addition to fusing with Golgi cisternae,
  vesicles also fuse with the ER with a rate that is controlled
  by the product of the concentrations of the vesicular ER v-SNARE and a
  fixed concentration of ER t-SNAREs. Due to the expansive volume of
  the ER that brings it in proximity to the entire Golgi apparatus,
  we assume that all vesicles can fuse with the ER independent of
  their originating cisterna. \\*
4. The age-dependent decrease (loss) of the cisternal concentration of
the $\a$ t-SNARE, which is essential for triggering the
  retrograde directionality of the traffic, occurs due to transport of
  vesicles to the ER. Therefore, no additional decay term is
  introduced for it. A small age-dependent decay term is introduced for the $\b$ t-SNARE (see
Eq.~(\ref{main_model})). 
\begin{figure}
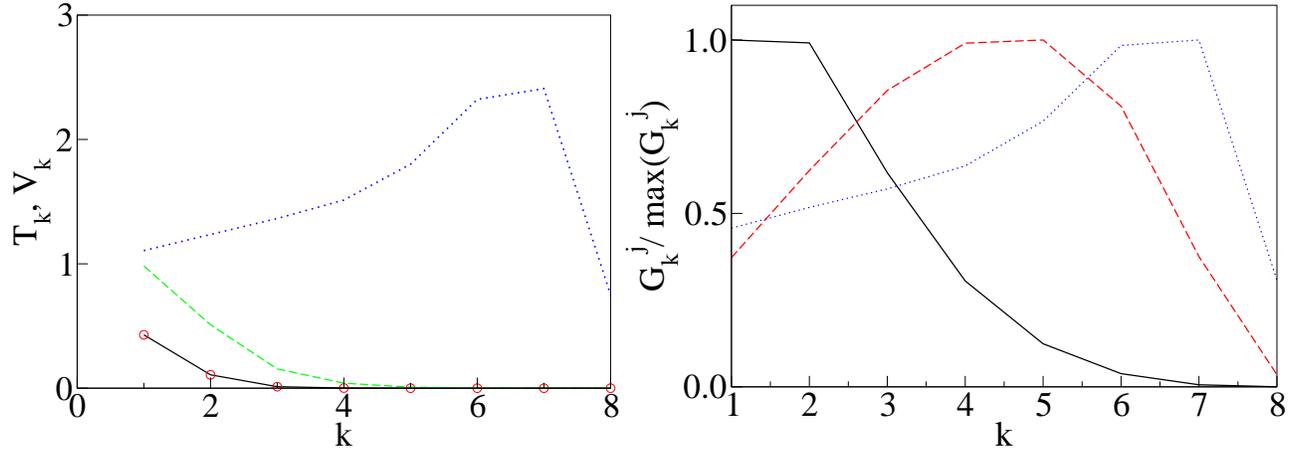

\begin{center}
\includegraphics[width=0.45 \textwidth]{f3a.eps}
\includegraphics[width=0.49 \textwidth]{f3b.eps}
\end{center}
\caption{\label{f3} {\bf Self-generated steady state distributions of alpha and beta SNAREs and enzymes as it is observed in the Golgi apparatus.}
{\bf Left panel:} $\a$ t-SNARE (solid line) and v-SNARE (circles) coinciding
with t-SNARE, and $\b$ t-SNARE (dashed  line) and v-SNARE
(dotted line) vs. the cisternal number $k$. 
{\bf Right panel:} Cis (solid line), medial (dashed line) and trans (dotted line) Golgi enzymes
  normalized by their maximum value vs number of cisterna $k$. The parameters are: Decay rates $\eta$
are zero for all substances except for $\b$ t-SNARE for which $\eta \tau=0.1$,
 the vesicular transport coefficient $\gamma \beta S B\tau=11$, and dissociation constants for
vesicular binding are $K=0.2$ for ER v-SNARE, $K=0.4$ for $\a$ t- and v-SNAREs, $K=1$ for $\b$ t-SNARE, $K=5$ for $\b$ v-SNARE, $K=1.4$ for cis enzymes, $K=2.5$ for median enzymes, and $K=5$ for trans enzymes. Initial concentrations of all substances in the first cisterna are $G_0^j=1, \;j=1,\ldots,8$, and the concentration of t-SNARE in ER is 0.7. }
\end{figure}
The dissociation constants for binding to vesicular sites are
summarized in the Table ~\ref{t1}. 
\begin{table}
    \begin{tabular}{|l|l|}
        \hline
         Substance          & Dissociation constant $K$ \\ \hline
        ER v-SNARE          & 0.2                       \\ 
        $\alpha$ t-SNARE    & 0.4                       \\ 
        $\alpha$ v-SNARE    & 0.4                       \\ 
        $\beta$ t-SNARE     & 1                         \\ 
        $\beta$ v-SNARE     & 5                         \\ 
        {\it Cis} enzyme    & 1.4                       \\ 
 	  Medial enzyme 	     & 2.5                       \\ 
        {\it Trans} enzyme  & 5                         \\
        \hline
    \end{tabular}
\caption{Dissociation constants for binding to vesicular sites that yield the plots depicted in Fig.~3. }
  \label{t1}
\end{table}
We found a good qualitative
agreement between our results and the experimentally observed
concentration profiles. With the proper choice of dissociation
constants (Table 1) $\b$ t-SNARE decay rate, and vesicular transport
intensity, the model functions in the following way: The strong
coat-binding affinities of  $\a$ and ER SNAREs effectively package them
into vesicles that bud from the younger {\it cis} cisternae. These vesicles
have a high probability to fuse with the ER due to a substantial
concentration of the ER SNAREs. These vesicles also recycle a good
fraction of strong-binding {\it cis} enzymes, and a part of medial enzymes
to the ER. The recycling of  $\a$ t-SNAREs to the ER seeds a cisternal
gradient, which is responsible for the mostly retrograde direction of
vesicular transport in the early cisternae. The recycling of $\b$
t-SNAREs to the ER is poor, yet when coupled with age-dependent decay,
the $\b$ t-SNAREs extends the {\it cis}$>${\it trans} t-SNARE gradient to the {\it trans}
Golgi. In more mature cisternae where the  $\a$ and ER SNAREs and
{\it cis}-enzymes are depleted, the vesicles incorporate the weaker-binding
molecules, such as medial and, to a lesser extent, {\it trans} enzymes and $\b$
SNAREs. These vesicles have a much lower probability to reach the ER
and transport their cargo mostly to younger Golgi cisternae. Finally,
the {\it trans}-most cisternae bud vesicles that contain predominantly {\it trans}
enzymes and $\b$ v-SNAREs. These cargoes are transported mostly
retrograde, but hardly reach the ER. The total fraction of each
protein that is retained in the Golgi (as compared to that recycled to
the ER) can be appreciated by its concentration in the {\it trans}-most
cisterna in Figs. ~\ref{f3}.  Since the
figures represent the situation before the last cisternal maturation
step and removal of the last cisternae, the protein concentration that
remains in the Golgi is equal to the initial concentration (set equal
to one for all molecules), minus the loss to the ER.

Our prediction that {\it cis}-enzymes,  $\a$ and ER SNAREs recycle through the
ER at a higher level than $\b$ SNAREs and {\it trans} enzymes is indeed born
out by numerous experimental observations in yeast and mammalian
cells. {\it Cis} but not {\it trans} Golgi markers accumulated in the ER upon an
acute ER-exit block  \cite{wooding1998dynamics}, \cite{jarvela2012irradiation}
or in the ER-derived intermediate compartment (ERGIC) after a
temperature-induced exit block from this compartment 
\cite{zhang2001ykt6}, \cite{xu2002gs15}.

Based on the SNARE dissociation constants that yielded the experimentally observed protein gradients (Table 1) we further predict that monomeric ERS24, which functions as ER-v-SNARE, has the highest affinity of all SNAREs for the COPI coat, followed by the $\a$ v- and t-SNAREs, (rBet1p and the proteins Syntaxin5 and Membrin, which together with ERS24 make up the $\a$ t-SNARE complex).  Indeed, ERS24 is much higher concentrated in COPI vesicles than any of the other v-SNAREs (Fig.~8B in \cite{volchuk2004countercurrent}). Syntaxin 5 is translated as a long and short version in mammalian cells \cite{hui1997isoform}. The longer form features a known ER-retrieval signal and we predict that it is this form that predominantly functions in the $\a$ t-SNARE complex and is more efficiently incorporated into COPI vesicles then the short form that likely functions mostly as $\b$ t-SNARE, which has a higher dissociation constant then the $\a$ t-SNARE. 

\begin{figure}
\begin{center}
\includegraphics[width=0.65 \textwidth]{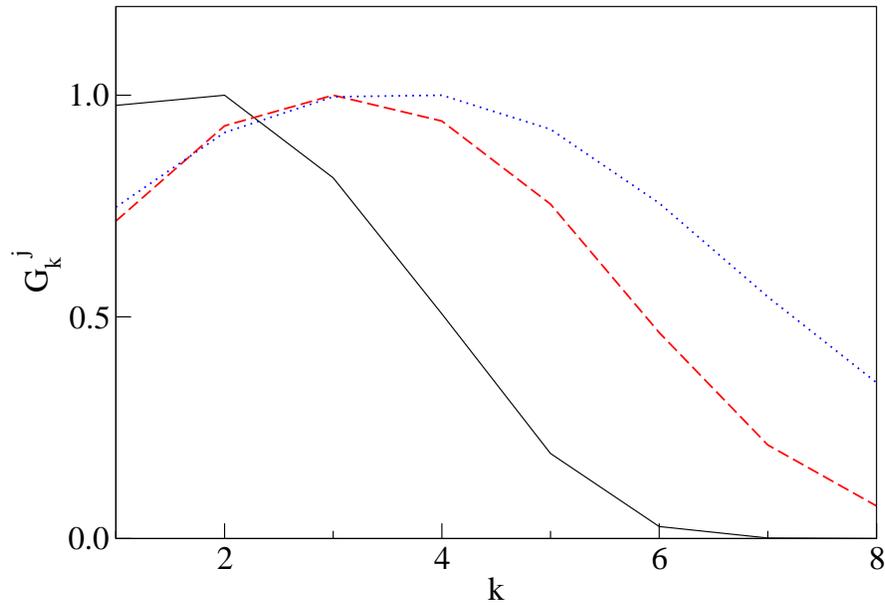}
\end{center}
\caption{\label{f4} {\bf Steady state concentrations of three
  classes of enzymes in the unrestricted fusion scenario.}  
Golgi enzyme concentrations are
  normalized by their maximum value. 
The parameters are: Decay rates for the {\it cis} and  {\it trans} t-SNARE are $\eta \tau=0.4$,
 the vesicular transport
coefficient $\gamma \beta S B\tau=4$, and dissociation constants for
vesicular binding are: $K=0.6$ for the ER v-SNARE, $K=0.4$ for the {\it cis} t- and
v-SNAREs, $K=1$ for {\it trans} t-SNARE, $K=5$ for {\it trans} v-SNARE, $K=0.6$
for {\it cis} enzymes, $K=2.5$ for medial enzymes, and $K=6$ for {\it trans}
enzymes. Initial concentrations of all proteins in the first
cisterna are $G_0^j=1, \;j=1,\ldots,8$, and the concentration of the
t-SNARE in the ER is 0.7. 
 }
\end{figure}

So far we assumed that vesicles only fuse with the immediate neighbors
of their progenitor cisternae. 
A stacked Golgi, however, is not a requirement for Golgi asymmetry and
cisternal maturation, which are also observed in {\it S. cerevisiae} where
individual Golgi cisternae are scattered throughout the cytoplasm
\cite{preuss1992characterization}, \cite{losev2006golgi},
\cite{matsuura2006live}, \cite{rivera2009rab}.
Removing the local
fusion restriction, and allowing vesicles to fuse with any cisterna and
the ER depending on their SNARE concentrations, we achieve only poor
enzyme segregation with all enzyme maxima shifted towards younger
cisternae, (Fig.~\ref{f4}).

We suggest therefore, that
a realistic description of fusion probability in {\it S. cerevisiae} must
include a factor that considers fusion preferences related to
cisternal age although it might be less stringent than the nearest
neighbor limitation of a Golgi stack. Golgi scattering occurs when
novel cisternae emerge from multiple, short-lived transitional ER
(tER) sites rather than from a single, stable tER \cite{papanikou2009yeast}.
If individual tER sites release multiple cisternae in
short succession before ceasing their activity, the diffusion limits
imposed by the ER-network could maintain sister cisternae that are
close in age in proximity to each other, thus {\it ad hoc} generating a
series of maturing Golgi cisternae that remain separate from those
generated in parallel by other tER sites.  Evidence from a recent
study by Nakano et al in {\it S. cerevisiae} supports this prediction:
When due to altered ER-morphology the motility of Golgi elements away
from the ER-exit site(s) is impeded, {\it cis} and {\it trans} Golgi
elements could be seen in close proximity to each other and to ER-exit
sites  \cite {okamoto2012high}. A position-age correlation is also
apparent from the more coarse-grain viewpoint: Consider the emission
of Golgi elements from multiple scattered ER exit sites and their
subsequent one-dimensional diffusion in the cytoplasmic half-space
away from ER membrane. The average distance from the ER membrane of a
Golgi element at time t after emission scales as $\sqrt{t}$. Thus, the
older Golgi elements are on average further away from the ER than the
younger ones.
Real-time imaging maps of
the spatial relationship between yeast Golgi cisternae that exchange
cargo should provide the experimental framework to make our model
applicable to Golgi systems with scattered compartments where we
expect the enzyme distribution to be somewhere in between the examples
shown in Fig.~\ref{f3} and Fig.~\ref{f4}.

\section{Discussion}
We present a simple model that explains the establishment and
maintenance of directed vesicular flow and concentration gradients in the Golgi apparatus, an organelle system that undergoes constant rejuvenation
by adding a new cisterna at the cargo-entering {\it cis} side while
dissolving the oldest cisterna as secretory and lysosomal cargo exit at the {\it trans} end.
Age is indeed the distinguishing feature of individual
Golgi cisternae that we identify as the key to symmetry breaking. As
cisternae mature the concentration of their functional SNAREs decreases, thereby providing the
seed for a {\it cis}$>${\it trans} cisternal gradient of fusion factors for
transport vesicles. This SNARE gradient causes the
predominantly retrograde direction of vesicular flux that retrieves
Golgi resident proteins, such as the SNAREs themselves and enzymes, from older to
younger cisternae and back to the ER. The vesicular transport of SNAREs  further
enhances their gradient until a steady state between the retrograde
vesicular and anterograde
cisternal progression is reached. Both the seed gradient and cisternal
maturation are indispensable for this outcome.  

The ``seeding'' temporal decay of cisternal SNARE concentrations
occurs via several mechanisms:
i) Retrieval to the ER
alone can 
account for the loss of the SNAREs present mostly in the young
cisternae.  However, the retrieval to the ER of the  Golgi SNAREs from
the medial and {\it trans} cisternae
is not  sufficient to create a seed gradient.  
ii) Experimental evidence for one such late Golgi SNARE,
Gos28, are compatible with the notion that its loss occurs through degradation: The levels of Gos28 can go up as
much as 40\%  when the availability of its chaperone GATE-16 is increased, preventing Gos28's proteolytic degradation \cite
 {sagiv2000gate}, \cite{zhong2011osbp}. 
 Gos28-levels also increase  when components of the Golgi-tethering complex COG are overexpressed
 \cite{oka2004cog}.
 This adjustability 
 means that a fraction of Gos28 is indeed wasted under
 the normal operational conditions.
iii)  Loss of SNAREs may also involve mechanisms in which
 Golgi-SNAREs become diverted to extra-Golgi functions. In yeast,
 Golgi-derived vesicles were shown to serve as source for
 autophagic membranes, which are later retrieved back to the Golgi
 \cite{ohashi2010membrane}, \cite{yamamoto2012atg9}.
The Gos28 homologue
 Gos1p in particular, has been implicated in the retrieval of the
autophagic membrane protein Atg9 to the Golgi
 \cite{ohashi2010membrane}.  iv) The loss of function of $\b$ t-SNARE in
 older cisternae may occur  due to modification of the membrane properties.
 v) A fraction of the decay of the late Golgi  t-SNARE is
due to its inactivation by  the corresponding 
v-SNARE with its emerging counter-current gradient (see Fig.~\ref{f3}).
Cognate SNARE complexes not only assemble when present on
opposite membranes (i.e. in {\it trans}) but also when present at the same
membrane (i.e. in {\it cis}), where most of them are disrupted under energy
expenditure by the NSF/$\a$SNAP machinery  \cite{lang2002snares},
\cite{xu2002gs15}. 
Nevertheless, in freshly isolated plasma membranes, where
the v-SNARE concentration is low, about 10\% of t-SNAREs are found in
unproductive SNARE complexes \cite{bar2008imaging}. 
As the
v-SNARE concentration goes up from {\it cis}-  to {\it trans}-Golgi
(blue line in left panel of Fig.~\ref{f3}) concomitant with
the decreasing t-SNARE levels (green line in left panel of
Fig.~\ref{f3}), binding of the t-SNARE into fusion-incompetent SNARE
complexes will  sharpen the  {\it cis}$>${\it trans} 
gradient of its fusion-competent concentration.

Once the retrograde vesicular flux is established, different affinities of Golgi
enzymes for the  vesicles  explain the
enzyme peaks in {\it cis}, medial and {\it trans} cisternae. One finding from our
simulations is that the differential distribution of Golgi proteins
can only be achieved when the vesicles are allowed to recycle back to
the ER. This is in good agreement with experimental
observations \cite{storrie1998recycling}, \cite{wooding1998dynamics}, \cite{jarvela2012irradiation}.
However, the importance of Golgi protein cycling
through the ER for the enzyme segregation
had not been appreciated in previous models that explained the Golgi
enzyme peaks  \cite{glick1997cisternal}, \cite{weiss2000protein}
because of the arbitrarily implementation of the
directionality of vesicle transport.
 It should be possible to test this important conclusion
  from our model experimentally. In yeast, ER- recycling of
  Golgi-derived vesicles can be stopped and the consequences for the
  segregation of {\it cis} and {\it trans} Golgi enzymes can be monitored by dual
  color time-lapse microscopy 
  \cite{losev2006golgi}, \cite{matsuura2006live}. 
 This approach is feasible in strains harbouring
  temperature-sensitive mutations in ER-t-SNAREs
  \cite{lewis1996snare}, \cite{lewis1997novel}.
  Importantly, the switch to the non-permissive temperature
  does not lead to the accumulation of Golgi-derived transport
  vesicles in these strains, presumably because ER-destined vesicles
  also contain significant amounts of $\alpha$ v-SNAREs, which allows
  them to efficiently fuse with the Golgi when fusion with the ER is
  thwarted. Such a scenario is indeed consistent with the SNARE
  dissociation constants of our model (Table 1).

Our simulations are insensitive to a broad spectrum of initial
conditions. Regardless of whether we started with a single cisterna
and added new cisternae one by one as it would occur during Golgi {\it
  de novo} formation, or considered a complete stack of identical
cisternae when turning on the vesicular flux and SNARE loss mechanism,
in each case the same steady state was reached.

An important question is the relevance and specificity of constants
used for the modeling . Naturally, the range of admissible constants narrows as one reproduces more detailed and specific scenarios. Our
first observation, that a temporal loss of SNAREs results in directed
vesicular flux, is very general and holds for virtually any set of
constants (see Eqs.  (\ref{lambda}, \ref{form3})).
The selection of constants became more
restrictive when the {\it cis}, medial, and {\it trans} peaks of Golgi enzymes and
the actual 2 SNARE pair scenario were reproduced. The dissociation
constants for binding of SNAREs and enzymes to vesicular sites had to
be tuned within a ~10\% precision. The actual values of the
dissociation constants
are of the same order as protein concentrations, which  is quite common for protein-protein interactions and appears to be
evolutionally attainable \cite{kumar2006pint}. Furthermore, to reproduce the shape of experimentally
measured enzyme and SNARE peaks, the directionality of vesicular flux needed to be
sufficiently strong, which we attempted to achieve while minimizing
the  decay term for SNAREs. The SNARE decay and vesicular transport
constants did not have to be tuned as precisely as the dissociation
constants and their admissible variation range is generally 20-30\%.


We observed that the distinct enzyme peaks can be achieved with just one
cognate SNARE pair. Why then does the Golgi afford two SNARE pairs?
One proposal, put forward by Volchuk et al., is that only the $\a$ SNARE
mediates retrograde transport of Golgi resident proteins while the $\b$
SNARE is dedicated to anterograde transport of exocytic and lysosomal
cargo \cite{volchuk2004countercurrent}. 
We consider this unlikely, however, based on the collective
experimental evidence. Immuno electron microscopy-based observations
of anterograde cargo in COPI vesicles is controversial and more recent
organelle proteomics readily identified Golgi resident proteins but no
exocytic cargo in COPI vesicles (reviewed in \cite{gannon2011golgi}).
Moreover, functional data in yeast
have provided unequivocal evidence for a role of the $\b$ SNARE in Golgi
enzyme trafficking. Thus, acute inhibition of the $\b$ SNARE Sft1 leads
to a rapid loss of {\it trans} and medial Golgi enzymes from Golgi cisternae
and their dispersion into vesicles that are apparently unable to
fuse \cite{wooding1998dynamics}.
Therefore, both SNARE pairs are likely to operate in tandem rather than
in a countercurrent fashion. 
Although the concentration of vesicular
SNAREs does not influence the directionality of fusion, it determines
fusion efficiency. Thus, high concentrations of one of each v-SNAREs on
either end of the Golgi can sustain efficient vesicular traffic
throughout the Golgi stack.
 In addition, each SNARE pair could have
distinct, additional roles at the Golgi boundaries. While this is well
established only for the $\a$ SNARE, which mediates fusion of ER-derived
vesicles at the {\it cis} face (reviewed in
\cite{malsam2011organization}),
recent evidence suggests that $\b$ SNAREs GS15
and Ykt6 can participate in the fusion of endosomes with the
{\it trans} Golgi or TGN \cite{tai2004participation}.

According to our model, the experimentally observed steep {\it
  cis}$>${\it trans}
gradient of the $\a$ SNARE results in an almost sequential action of the
two SNAREs within the maturing Golgi stack. This in turn yields two
{\it de
facto} distinct COPI vesicle populations, one enriched in $\a$ SNAREs and
{\it cis} Golgi markers, the other in $\b$ SNAREs and enzymes from the medial
and {\it trans} Golgi. Plant Golgi stacks indeed feature morphologically
distinct vesicles around the rim of the {\it trans} and {\it cis} cisternae,
respectively  \cite{staehelin2008nanoscale}
and in mammalian cells COPI vesicles enriched in either
{\it cis} and {\it trans} Golgi proteins and the corresponding SNAREs have been
distinguished \cite{lanoix2001sorting}, \cite{malsam2005golgin}, \cite{martinez2007low}.
In our model these two subpopulations of COPI vesicles
are simply due to differences in the competitiveness of the SNAREs and
enzymes for binding to a universal COPI-coat rather than to two
vesicle types that differ in the composition of the COPI coat or, more
generally, in the machinery for cargo selection. 
Even though vertebrates have been reported to  possess  several  COPI isoforms
 \cite{moelleken2007differential}, we show that a single COPI species,
 as in fungi and plants, is sufficient
generate the variance in vesicle content.

In summary, we have presented an explanation for why the minimal
requirement of one SNARE pair and one vesicle type for the generation
and maintenance of each distinct organelle
\cite{heinrich2005generation}
is relaxed for organelles that evolve from each other through maturation.
Apart from the Golgi apparatus this might also be relevant for the
organelles along the plasma membrane-early endosome-late endosome
axis.

\section{Methods}
\subsection{Establishment of a {\it cis}$>${\it trans} SNARE gradient that mediates
 retrograde vesicular flow}
Here we do not specify the nature of t- and v- SNAREs, 
simply calling  fusiogenic molecules present in a vesicle and cisterna
v-SNAREs and t-SNAREs.
The chemical distinction between t- and v-SNAREs will be stated
later. However, to keep the same notations
throughout the paper, we use the specific $v, V$ and $t, T$
notations already here. 
Small
letters denote the vesicular concentrations of a molecule with the
subscript referring to the parental cisterna. So $v_k$ and $t_k$ are
concentrations of v- and t-SNAREs, and $g_k^J$ is the
concentration of the $j$th Golgi enzyme in a vesicle that emerged from
the $k$th cisterna . Capital letters $V_k$, $T_k$, and $G_k^j$  denote
the concentrations of these substances in cisterna number $k$. 
We number the compartments in the {\it
  cis} to {\it trans} direction, so the youngest cisterna has number
one. 

The number of vesicles that bud from the $k$th compartment per unit
time,  $dN_k/dt|_{bud}$, is assumed to be proportional to the
area of the compartment $S_k$,
\begin{equation}
\label{bud}
\left.\frac {dN_k}{dt}\right|_{bud}=\beta S_k,
\end{equation}
where $\beta$ is the budding rate constant which depends on the
concentration and activity of coat proteins.

A vesicle emitted from the $k$th cisterna fuses with the $i$th
cisterna with a probability proportional to the product of the
concentrations of the SNARE in the vesicle $v_k$ and the SNARE in the cisterna
$T_i$. The number of vesicles that fuse with the cisterna $i$ per unit
time is 
\begin{equation}
\label{fuse}
\left.\frac {dN_k}{dt}\right|_{fuse\;to\; i}=\alpha N_k v_k T_i. 
\end{equation}
with  $\alpha$
being the fusion rate constant.   
The assumption of local transport restricts  a vesicle
emitted by the $k$th cisterna to
fuse with the $(k-1)$th, $k$th, and $(k+1)$th cisternae.
The {\it trans}-most cisterna does not receive any retrograde vesicular
input. The time evolution of the population of vesicles emitted by the $k$th
compartment is described by the rate equation which includes both the
budding and fusion terms. 
\begin{equation}
\label{total}
\frac {dN_k}{dt}=\beta S_k - \alpha N_k v_k (T_{k-1}+T_k + T_{k+1}). 
\end{equation}
At steady state, the concentration of vesicles emitted by
the $k$th compartment becomes
\begin{equation}
\label{steady}
N_k^*=\frac{\beta S_k}{\alpha v_k(T_{k-1}+T_k+T_{k+1})}.
\end{equation}
Hence, an increment in SNARE concentration in the $k-1$st cisterna
due to the  vesicular flux from the $k$th
cisternae is
\begin{equation}
\label{flux}
\left. \frac {d T_{k-1}}{dt}\right|_{k-> k-1}=\beta S_k \gamma t_k \frac{T_{k-1}}{(T_{k-1}+T_k+T_{k+1})}.
\end{equation}
 A dimensionless factor $\gamma$  describes how the cargo is
``diluted'' when a vesicle fuses with its target cisterna and is equal to the ratio of vesicle
to compartment surface areas. 
Assuming mass-action equilibrium between the vesicular binding sites
and its cargo (t-SNARE) and that budding of a single vesicle does not
significantly alter the cisternal concentration of t-SNARE, the amount of t-SNARE in a vesicle is
$$
t_k=\frac{B T_k}{T_k+K}.
$$
Here $B$ is the concentration of cargo binding sites in a vesicle
and $K$ is the dissociation constant for binding between cargo and
such sites.
Eq. (\ref{flux}) indicates that
the steady state flux of vesicles does not depend on the v-SNARE
concentration and is only determined by the budding rate and t-SNARE
distribution.  In the following we assume that the
volume and the budding area of the compartments remains constant,
$S_k=S$.  Relaxing this assumption does not substantially change the results. 

The rate equation that describes  the evolution of the t- SNARE
concentration in the $k$th compartment reads
\begin{align}
\nonumber
\frac {d T_{k}}{dt}= - \eta  T_{k}  + \gamma \beta  S \times\\
\label{main}
\times\left[ - t_k\frac{T_{k-1}+T_{k+1}}{T_{k-1}+T_{k}+T_{k+1}}
\right.+\\
\nonumber
+ t_{k-1} \frac {T_{k}}{T_{k-2}+T_{k-1}+T_{k}} +\\
\nonumber
+\left. t_{k+1}  \frac{T_{k}}{T_{k}+T_{k+1}+T_{k+2}}
\right].
\end{align}
The first term describes the loss of
the t-SNARE with the per molecule rate $\eta$. To keep it general, the
loss term collects all
mechanisms of t-SNARE decay approximately described by first-order kinetics, such
as degradation, loss of mis-targeted vesicles, etc. Thus, here  the
lost vesicles are not accounted for in Eqs.
(\ref{total}, \ref{steady}), but are only included in the first term in
  Eq. (\ref{main}). 
The second line 
describes the outgoing vesicular transport from the $k$th cisterna to
its $(k-1)$th and $(k+1)$th neighbors, and the last two lines
represents the incoming flux from the same neighbors to the $k$th
cisterna. 

To complete the description of  t-SNARE distribution, the vesicular
transport equation (\ref{main}) has to be complemented by the
definition of cisternal dynamics: Every $\tau$ time units the running
number of each cisterna is incremented by one, so that the $k$th
cisterna becomes $k+1$st. A new first cisterna with a given initial
concentration of t-SNARE $T_0$ 
is added to the {\it cis} end of the stack,
while the {\it trans}-most cisterna is removed.  

We measure cisternal concentrations of SNAREs and other
molecules in the units of their initial concentrations in the first cisterna and
the natural unit of time is the period of cisternal maturation $\tau$.
This is equivalent to setting these quantities equal to one. 
Then the t-SNARE distribution is described by three parameters:
decay rate $\eta$, the vesicular transport coefficient $\gamma \beta
S B$, and the dissociation constant $K$. 

This cisternal maturation scenario together with
Eq. (\ref{main}) are implemented numerically as a simple Euler update. 
For reasonable values of parameters the
system quickly converges to a steady  regime: In each cisterna
concentrations of t- and v-SNAREs undergo periodic evolution with the
period $\tau$.  Plots of the
cisternal distributions of the t-SNARE are presented  in Fig.~\ref{f2} in the
Results. 

\subsection{Analytic solution for the asymptotic steady state cisternal
  concentrations of SNAREs}
Consider a hypothetical system where the number of cisternae is
non-biologically large. 
For older cisternae, the concentrations of
SNAREs are small, $T_k \ll K$,  so the uptake of a SNARE into a
vesicle is proportional to the concentration of the SNARE in the
progenitor cisterna, $t_k=T_k B/K$. 
In the asymptotic regime, i.e., sufficiently far from the first and the
last cisterna, we seek a solution of Eq.~(\ref{main}) in the form
\begin{equation}
\label{form}
T_k(t)=\psi (t)\lambda^K.
\end{equation}
After substitution into (\ref{main}) it yields
\begin{equation}
\label{form2}
T_k(t)=T_0 \; \lambda^K \exp[-f(\lambda)t],
\end{equation}
where 
\begin{equation}
\label{f}
f(\lambda)=\eta + \frac{\gamma \beta  S B}{K} \; \frac{(\lambda-1)^2}{\lambda^2 +
    \lambda +1}.
\end{equation}
We look for the periodic
solution in a sense that each $\tau$ time units, after the addition
of a new cisterna and dissolution of the most mature cisterna, the system returns to
the same state. So  the $k$th cisterna  at the time $t+\tau$ must be
identical to the $k+1$ cisterna at the time $t$,
\begin{equation}
\label{steadys}
T_{k+1}(t) = T_k(t+\tau).
\end{equation}
This yields the following equation for $\lambda$,
\begin{equation}
\label{lambda}
\ln(\lambda)=-\tau\eta - \frac{\tau\gamma \beta  S B}{K} \; \frac{(\lambda-1)^2}{\lambda^2 +
    \lambda +1}
\end{equation}
which is solved numerically. 

We observe that in the asymptotic regime, the steady state t-SNARE
concentration decays exponentially with the number of
cisterna,  
\begin{equation}
\label{form3}
T_k^{steady} =T_0 \; \lambda^k(\tau\eta, \tau\gamma \beta S B /K) 
\end{equation}
with the coefficient $\lambda(\tau\eta, \tau\gamma \beta S B /K)$ being the solution of Eq. (\ref{lambda}).

\begin{figure}
\begin{center}
\includegraphics[width= 0.65 \textwidth]{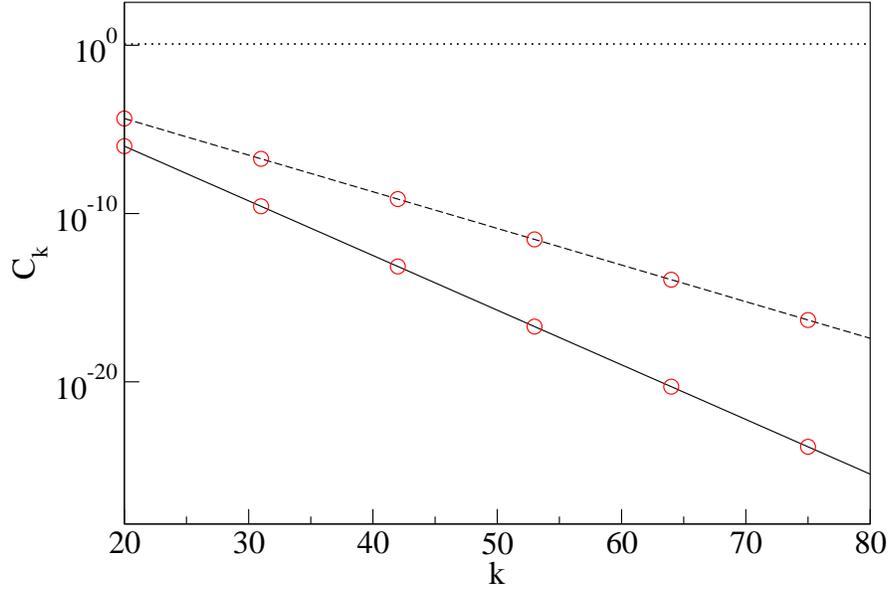}
\end{center}
\caption{\label{f5}
{\bf  Distribution of SNARE for large number of cisternae.} The steady state
  gradient has the exponentially decaying form,  $T_k=T_0 \lambda^k$
  where $\lambda$  depends on two dimensionless groups of parameters,
  $\eta\tau$ which caracterizes the decay of the SNARE and 
$\gamma  \beta \tau S B /K$ which characterizes the vesicular transport of SNARE:  $\eta \tau=0.5$ and $\gamma \beta \tau S B/K =1.5$ (solid line) with the best fit given by 
$\lambda \approx 0.606$,  $\eta \tau=0.5$
  and $\gamma  \beta \tau S B/ K =0$ (dashed line) with the best fit given by 
$\lambda\approx 0.476$ .
Theoretical values of $\lambda$ determined from (\ref{lambda}) are
indistinguishable from the values obtained as  the best fit for
simulations. 
Dotted line corresponds to the
case of zero loss, $\eta \tau=0$
  and $\gamma \beta \tau S B/ K =1$, and illustrates the absence of a concentration gradient. To reveal the exponential decay of SNARE concentration,
  we purposefully consider a non-biologically high number of
  compartments and analyze the SNARE concentration away from both the
  {\it cis} and {\it trans} ends of the stack, where the boundary effects can play
  a role. 
}
\end{figure}

Simulations confirm our theoretical prediction
given by
(\ref{form}, \ref{lambda}), see Fig.~\ref{f5}.

The necessity of the loss term for establishing the gradient by breaking the initial symmetry between the
cisternae is clearly revealed by the following analytic
argument: For a small loss rate ($\eta \tau \ll 1$), the expansion of the steady
state exponent $\l$ reads
\begin{equation}
\label{small_loss}
\lambda=1-\eta\tau - \frac{\gamma \beta \tau S B }{K} \frac{(\eta\tau)^2}{3}
+\mathcal O (\eta\tau)^3.
\end{equation}
Hence $\lambda=1$ for $\eta\tau=0$ independent of the intensity of
the vesicular transport characterized by $ \gamma \beta \tau S B/K$.
Indeed,
without breaking the initial similarity between cisternae, a vesicle would fuse with any of its three
target compartments with the same probability, so that vesicular flux
into a compartment would be equal to the vesicular flux
out of this compartment.  In other words, the vesicular transport can
only enhance the initial difference in concentrations between
cisternae, created by some other mechanism, rather than create this
difference {\it de novo}.

\subsection{Establishment of Golgi enzyme peaks in {\it cis}, medial and {\it trans} cisternae via SNARE-mediated retrograde vesicular traffic}
The transport of Golgi enzymes with cisternal concentrations $G^j$,
where $j=1,\ldots,3$ labels an enzyme class, is described by an equation
analogous to  Eq.~(\ref{main}). 
The difference is that instead of  a single vesicular cargo type (t-SNARE), we now consider three
classes of competitors for vesicular seats. 
Thus, for each $j$, $ {d G_{k}^j}/{dt}$ replaces the $ {d T_{k}}/{dt}$
in the
left-hand side and $g_k^j$ replaces $t_k$ in the right-hand side of
Eq.~(\ref{main}), 
\begin{align}
\nonumber
\frac {d G_{k}^j}{dt}= - \eta_j  G_{k}^j  + \gamma \beta  S \times\\
\label{main_cargo}
\times\left[ - g_k^j\frac{T_{k-1}+T_{k+1}}{T_{k-1}+T_{k}+T_{k+1}}
\right.+\\
\nonumber
+ g_{k-1}^j \frac {T_{k}}{T_{k-2}+T_{k-1}+T_{k}} +\\
\nonumber
+\left. g_{k+1}^j  \frac{T_{k}}{T_{k}+T_{k+1}+T_{k+2}}
\right].
\end{align}
Here (only in this subsection)  we assume that the enzyme transport does
not affect the vesicular flow, which is established by the
autonomously evolving t-SNARE
distribution, described
by  (\ref{main}).  The concentration 
of enzyme  $g_k^j $ uploaded to a vesicle is determined by
the mass action equilibrium
\begin{align}
\label{cargo2}
g_k^j K_j = B'   G_k^j, \; j=1,\ldots,3\\
\nonumber
B=B'+\sum_{j=1}^3 g_k^j
\end{align}
Here $K_J$ are vesicle-$j$th enzyme  dissociation constants, the last equation states
that the total number of the vesicular binding sites $B$ is equal to
the number of free sites $B'$ plus the number of sites occupied by
enzymes of all three classes.  
Solving (\ref{cargo2}), one finds $g_k^j$,
\begin {equation}
\label{cargo_v}
g_k^j=\frac{BG_k^j}{K_j(1+\sum_{i=1}^3 G_k^i/K_i)},
\end{equation} 
which are subsequently substituted into Eq.~(\ref{main_cargo}), 
As with SNAREs, each newly formed (first) cisterna is assumed to be
loaded with Golgi enzymes with given concentrations, $G_1^j(t=0)=G_0^j$
We assume that there is no temporal decay of enzymes, so $\eta_j$ is
put equal to zero in the transport equations.  

\begin{figure}
\begin{center}
\includegraphics[width= 0.65 \textwidth]{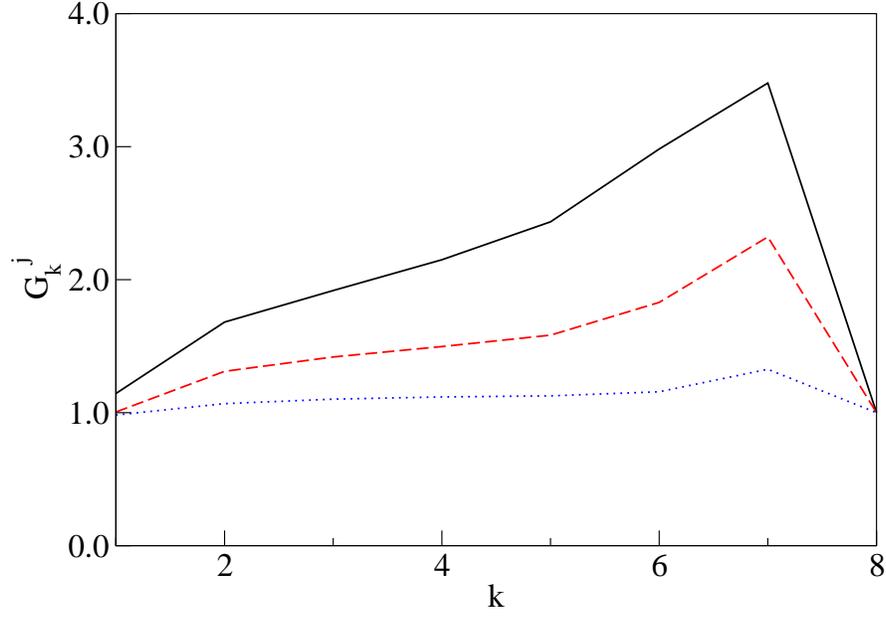}
\end{center}
\caption{\label{f6} {\bf Enzyme segregation depends on the open
    boundary condition.} Steady state 
  distribution of the same enzymes as in Fig.~\ref{f2} if the ``Closed boundary
  conditions'' on the {\it cis} end of the stack are assumed: No vesicles
  can exit the Golgi.  The parameters are the same as in
  Fig.~\ref{f2}.  }
\end{figure}

When the retrograde vesicular transport is counterbalanced by the
anterograde cisternal progression, the enzyme distribution reaches its
steady state. The nature of the steady state depends on the boundary
conditions imposed on the {\it cis} side of the Golgi stack: An ``open''
boundary condition is implemented as a zeroth cisterna (ER) with a
fixed concentration of t-SNAREs which can fuse vesicles (see Fig.~\ref{f2}, right
panel), while under the
``closed'' boundary conditions vesicles do not escape the Golgi, see Fig.~\ref{f6}). 

\subsection{The two Golgi SNARE pairs can function with a single vesicle type to establish their own gradients and the observed Golgi enzyme peaks in {\it cis}, medial and {\it trans} }
Putting together two mechanisms considered above, we introduce a
realistic model of Golgi transport. It describes the evolution of 8
distinct types of molecules:  $\a$ and $\b$ sets of t- and v-SNAREs controlling
intra-Golgi fusion, a v-SNARE for fusion with the ER,  and {\it cis},
medial, and {\it trans} types of enzymes.  For brevity of equations, we use
the universal notations $G_k^j$ and $g_k^J$ for cisternal and
vesicular concentrations of each of the eight molecules,
$j=1,\ldots,8$. At the same time, in the fusion rate terms we retain the
specific notations for t- and v-SNAREs with 
 superscripts ``$\alpha$'', ``$\beta$'' and ``ER'' denoting the
affiliation of particular SNAREs. 
The evolution of the cisternal concentration $G_k^j$, $j=1,\ldots,8$ of
each type of molecule is
described by the rate equation similar to (\ref{main_cargo}) with two
important distinctions. First, the rate of fusion of a vesicle with a
cisterna, previously given  by (\ref{fuse}), is now proportional to
the sum of the products of  the concentrations
of {\it cis} and {\it trans} SNAREs \cite{heinrich2005generation}.
\begin{equation}
\label{fuse2}
\left.\frac {dN_k}{dt}\right|_{fuse\;to\; i}=\alpha N_k (v_k^{\a}
T_i^{\a}+ v_k^{\b}T_i^{\b}).
\end{equation}
The increment in the vesicular cargo concentration in the $k-1$th cisterna 
due to the  vesicular flux from the $k$th
cisternae is (compare to (\ref{flux}),
\begin{equation}
\label{flux2}
\left. \frac {d G_{k-1}^j}{dt}\right|_{k-> k-1}=\beta S \gamma g_k^j \frac{v_k^{\a} T_{k-1}^{\a} + v_k^{\b} T_{k-1}^{\b}}
{\sum_{j=k-1}^{k+1} (v_k^{\a} T_{j}^{\a} + v_k^{\b}
  T_{j}^{\b}) + v_k^{ER} T_{ER}}.
\end{equation}
Here $g_k^j$ is the vesicular concentration of molecule $j$ defined by mass-action equilibrium (\ref{cargo_v}) between
vesicular binding sites and all eight competing molecules.
The last term in the denominator corresponds to the fusion of vesicles
with the ER, which is the second distinction of the considered mechanism
from the model case analyzed above.  The ER t-SNARE
concentrations $T_{ER}$ is considered to remain constant and vesicles originating
from any cisterna can fuse with the ER. Assembling together all gain and
loss mechanisms for the cisternal concentration of $G_k^J$, we write
the complete system of kinetic equations that describe the vesicular transport.
\begin{align}
\nonumber\
\frac {d G_{k}^j}{dt}=- \eta_j  G_{k}^j  + \gamma \beta  S \times\\
\label{main_model}
\times\left[ - g_k^j\frac{v_k^{\a} T_{k-1}^{\a} + v_k^{\b} T_{k-1}^{\b}+
v_k^{\a} T_{k+1}^{\a} + v_k^{\b}T_{k+1}^{\b}
}{\sum_{m=k-1}^{k+1} (v_k^{\a} T_{m}^{\a} + v_k^{\b}
  T_{m}^{\b}) + v_k^{ER} T_{ER}} \right.+\\
\nonumber
+ g_{k-1}^j \frac{v_{k-1}^{\a} T_{k}^{\a} + v_{k-1}^{\b} T_{k}^{\b}
}{\sum_{m=k-2}^{k} (v_{k-1}^{\a} T_{m}^{\a} + v_{k-1}^{\b}
  T_{m}^{\b}) + v_{k-1}^{ER} T_{ER}}+\\
\nonumber
+\left. g_{k+1}^j \frac{v_{k+1}^{\a} T_{k}^{\a} + v_{k+1}^{\b} T_{k}^{\b}
}{\sum_{m=k}^{k+2} (v_{k+1}^{\a} T_{m}^{\a} + v_{k+1}^{\b}
  T_{m}^{\b}) + v_{k+1}^{ER} T_{ER}}\right].
\end{align}

 The escape of a fraction of vesicles from the Golgi to the ER provides one part of a loss
mechanism necessary for seeding the gradient of t-SNAREs. Yet we do not exclude the possibility of  other
mechanisms of t-SNARE decay, so the $- \eta_j  G_{k}^j$ remains present in
the rate equation. In the simulations,
we set  $\eta$ for the $\b$ t-SNARE equal to a small value, while the
decay coefficients for all other substances are put equal to
zero.  
\subsection*{ Vesicular transport without nearest neighbor fusion restriction}
To model the vesicular transport in yeast, we used an equation similar
to Eq.~(\ref{main_model}) where the restriction of nearest neighbor
fusion was relaxed,
\begin{align}
\frac {d G_{k}^j}{dt}=- \eta_j  G_{k}^j  + \gamma \beta  S \times\\
\nonumber
\label{main_yeast}
\times\left[ - g_k^j\frac{\sum_{i=1}^n (v_k^{\a}  T_{i}^{\a} + v_k^{\b} T_{i}^{\b})
}{\sum_{i=1}^{n} (v_k^{\a} T_{i}^{\a} + v_k^{\b}
  T_{i}^{\b}) + v_k^{ER} T_{ER}} \right.+\\
+ \sum_{i=1}^n g_{i}^j \frac{  v_{i}^{\a} T_{k}^{\a} + v_{i}^{\b} T_{k}^{\b}
}{\sum_{m=1}^{n} (v_{i}^{\a} T_{m}^{\a} + v_{i}^{\b}
  T_{m}^{\b}) + v_{i}^{ER} T_{ER}}.
\end{align}
A typical steady state distribution of enzymes produced by the
unrestricted vesicular fusion is shown in Fig.~\ref{f4}.

\section*{Acknowledgments}
We acknowledge the contribution of T. Rapoport
who suggested this study and critically commented on the
manuscript. A.M. also thanks her lab for helpful discussions.  
This
project was supported by NIH/NIDDK R01KD064842-07A1 and FONDECYT,
Chile, grant 1110288.


\begin{thebibliography}{54}
\expandafter\ifx\csname natexlab\endcsname\relax\def\natexlab#1{#1}\fi
\expandafter\ifx\csname bibnamefont\endcsname\relax
  \def\bibnamefont#1{#1}\fi
\expandafter\ifx\csname bibfnamefont\endcsname\relax
  \def\bibfnamefont#1{#1}\fi
\expandafter\ifx\csname citenamefont\endcsname\relax
  \def\citenamefont#1{#1}\fi
\expandafter\ifx\csname url\endcsname\relax
  \def\url#1{\texttt{#1}}\fi
\expandafter\ifx\csname urlprefix\endcsname\relax\def\urlprefix{URL }\fi
\providecommand{\bibinfo}[2]{#2}
\providecommand{\eprint}[2][]{\url{#2}}

\bibitem[{\citenamefont{Farquhar and Palade}(1981)}]{farquhar1981golgi}
\bibinfo{author}{\bibfnamefont{M.}~\bibnamefont{Farquhar}} \bibnamefont{and}
  \bibinfo{author}{\bibfnamefont{G.}~\bibnamefont{Palade}},
  \bibinfo{journal}{The Journal of Cell Biology} \textbf{\bibinfo{volume}{91}},
  \bibinfo{pages}{77s} (\bibinfo{year}{1981}).

\bibitem[{\citenamefont{Volchuk et~al.}(2004)\citenamefont{Volchuk, Ravazzola,
  Perrelet, Eng, Di~Liberto, Varlamov, Fukasawa, Engel, S{\"o}llner, Rothman
  et~al.}}]{volchuk2004countercurrent}
\bibinfo{author}{\bibfnamefont{A.}~\bibnamefont{Volchuk}},
  \bibinfo{author}{\bibfnamefont{M.}~\bibnamefont{Ravazzola}},
  \bibinfo{author}{\bibfnamefont{A.}~\bibnamefont{Perrelet}},
  \bibinfo{author}{\bibfnamefont{W.}~\bibnamefont{Eng}},
  \bibinfo{author}{\bibfnamefont{M.}~\bibnamefont{Di~Liberto}},
  \bibinfo{author}{\bibfnamefont{O.}~\bibnamefont{Varlamov}},
  \bibinfo{author}{\bibfnamefont{M.}~\bibnamefont{Fukasawa}},
  \bibinfo{author}{\bibfnamefont{T.}~\bibnamefont{Engel}},
  \bibinfo{author}{\bibfnamefont{T.}~\bibnamefont{S{\"o}llner}},
  \bibinfo{author}{\bibfnamefont{J.}~\bibnamefont{Rothman}},
  \bibnamefont{et~al.}, \bibinfo{journal}{Molecular Biology of the Cell}
  \textbf{\bibinfo{volume}{15}}, \bibinfo{pages}{1506} (\bibinfo{year}{2004}).

\bibitem[{\citenamefont{Roth}(2002)}]{roth2002protein}
\bibinfo{author}{\bibfnamefont{J.}~\bibnamefont{Roth}},
  \bibinfo{journal}{Chemical Reviews} \textbf{\bibinfo{volume}{102}},
  \bibinfo{pages}{285} (\bibinfo{year}{2002}).

\bibitem[{\citenamefont{Grasse et~al.}(1957)}]{grasse1957ultrastructure}
\bibinfo{author}{\bibfnamefont{P.}~\bibnamefont{Grasse}} \bibnamefont{et~al.},
  \bibinfo{journal}{Comptes Rendus Hebdomadaires des Seances de l'Academie des
  Sciences} \textbf{\bibinfo{volume}{245}}, \bibinfo{pages}{1278}
  (\bibinfo{year}{1957}).

\bibitem[{\citenamefont{Glick and Luini}(2011)}]{glick2011models}
\bibinfo{author}{\bibfnamefont{B.}~\bibnamefont{Glick}} \bibnamefont{and}
  \bibinfo{author}{\bibfnamefont{A.}~\bibnamefont{Luini}},
  \bibinfo{journal}{Cold Spring Harbor Perspectives in Biology}
  \textbf{\bibinfo{volume}{3}} (\bibinfo{year}{2011}).

\bibitem[{\citenamefont{Glick and Malhotra}(1998)}]{glick1998curious}
\bibinfo{author}{\bibfnamefont{B.}~\bibnamefont{Glick}} \bibnamefont{and}
  \bibinfo{author}{\bibfnamefont{V.}~\bibnamefont{Malhotra}},
  \bibinfo{journal}{Cell} \textbf{\bibinfo{volume}{95}}, \bibinfo{pages}{883}
  (\bibinfo{year}{1998}).

\bibitem[{\citenamefont{Pelham}(1998)}]{pelham1998getting}
\bibinfo{author}{\bibfnamefont{H.}~\bibnamefont{Pelham}},
  \bibinfo{journal}{Trends in Cell Biology} \textbf{\bibinfo{volume}{8}},
  \bibinfo{pages}{45} (\bibinfo{year}{1998}).

\bibitem[{\citenamefont{Mart{\'\i}nez-Men{\'a}rguez
  et~al.}(2001)\citenamefont{Mart{\'\i}nez-Men{\'a}rguez, Prekeris, Oorschot,
  Scheller, Slot, Geuze, and Klumperman}}]{martinez2001peri}
\bibinfo{author}{\bibfnamefont{J.}~\bibnamefont{Mart{\'\i}nez-Men{\'a}rguez}},
  \bibinfo{author}{\bibfnamefont{R.}~\bibnamefont{Prekeris}},
  \bibinfo{author}{\bibfnamefont{V.}~\bibnamefont{Oorschot}},
  \bibinfo{author}{\bibfnamefont{R.}~\bibnamefont{Scheller}},
  \bibinfo{author}{\bibfnamefont{J.}~\bibnamefont{Slot}},
  \bibinfo{author}{\bibfnamefont{H.}~\bibnamefont{Geuze}}, \bibnamefont{and}
  \bibinfo{author}{\bibfnamefont{J.}~\bibnamefont{Klumperman}},
  \bibinfo{journal}{The Journal of Cell Biology}
  \textbf{\bibinfo{volume}{155}}, \bibinfo{pages}{1213} (\bibinfo{year}{2001}).

\bibitem[{\citenamefont{Malsam et~al.}(2005)\citenamefont{Malsam, Satoh,
  Pelletier, and Warren}}]{malsam2005golgin}
\bibinfo{author}{\bibfnamefont{J.}~\bibnamefont{Malsam}},
  \bibinfo{author}{\bibfnamefont{A.}~\bibnamefont{Satoh}},
  \bibinfo{author}{\bibfnamefont{L.}~\bibnamefont{Pelletier}},
  \bibnamefont{and} \bibinfo{author}{\bibfnamefont{G.}~\bibnamefont{Warren}},
  \bibinfo{journal}{Science} \textbf{\bibinfo{volume}{307}},
  \bibinfo{pages}{1095} (\bibinfo{year}{2005}).

\bibitem[{\citenamefont{Gilchrist et~al.}(2006)\citenamefont{Gilchrist, Au,
  Hiding, Bell, Fernandez-Rodriguez, Lesimple, Nagaya, Roy, Gosline, Hallett
  et~al.}}]{gilchrist2006quantitative}
\bibinfo{author}{\bibfnamefont{A.}~\bibnamefont{Gilchrist}},
  \bibinfo{author}{\bibfnamefont{C.}~\bibnamefont{Au}},
  \bibinfo{author}{\bibfnamefont{J.}~\bibnamefont{Hiding}},
  \bibinfo{author}{\bibfnamefont{A.}~\bibnamefont{Bell}},
  \bibinfo{author}{\bibfnamefont{J.}~\bibnamefont{Fernandez-Rodriguez}},
  \bibinfo{author}{\bibfnamefont{S.}~\bibnamefont{Lesimple}},
  \bibinfo{author}{\bibfnamefont{H.}~\bibnamefont{Nagaya}},
  \bibinfo{author}{\bibfnamefont{L.}~\bibnamefont{Roy}},
  \bibinfo{author}{\bibfnamefont{S.}~\bibnamefont{Gosline}},
  \bibinfo{author}{\bibfnamefont{M.}~\bibnamefont{Hallett}},
  \bibnamefont{et~al.}, \bibinfo{journal}{Cell} \textbf{\bibinfo{volume}{127}},
  \bibinfo{pages}{1265} (\bibinfo{year}{2006}).

\bibitem[{\citenamefont{Glick et~al.}(1997)\citenamefont{Glick, Elston, and
  Oster}}]{glick1997cisternal}
\bibinfo{author}{\bibfnamefont{B.}~\bibnamefont{Glick}},
  \bibinfo{author}{\bibfnamefont{T.}~\bibnamefont{Elston}}, \bibnamefont{and}
  \bibinfo{author}{\bibfnamefont{G.}~\bibnamefont{Oster}},
  \bibinfo{journal}{FEBS Letters} \textbf{\bibinfo{volume}{414}},
  \bibinfo{pages}{177} (\bibinfo{year}{1997}).

\bibitem[{\citenamefont{S{\"o}llner}(1995)}]{sollner1995snares}
\bibinfo{author}{\bibfnamefont{T.}~\bibnamefont{S{\"o}llner}},
  \bibinfo{journal}{FEBS Letters} \textbf{\bibinfo{volume}{369}},
  \bibinfo{pages}{80} (\bibinfo{year}{1995}).

\bibitem[{\citenamefont{Jahn and Scheller}(2006)}]{jahn2006snares}
\bibinfo{author}{\bibfnamefont{R.}~\bibnamefont{Jahn}} \bibnamefont{and}
  \bibinfo{author}{\bibfnamefont{R.}~\bibnamefont{Scheller}},
  \bibinfo{journal}{Nature Reviews Molecular Cell Biology}
  \textbf{\bibinfo{volume}{7}}, \bibinfo{pages}{631} (\bibinfo{year}{2006}).

\bibitem[{\citenamefont{Fasshauer et~al.}(1998)\citenamefont{Fasshauer, Sutton,
  Brunger, and Jahn}}]{fasshauer1998conserved}
\bibinfo{author}{\bibfnamefont{D.}~\bibnamefont{Fasshauer}},
  \bibinfo{author}{\bibfnamefont{R.}~\bibnamefont{Sutton}},
  \bibinfo{author}{\bibfnamefont{A.}~\bibnamefont{Brunger}}, \bibnamefont{and}
  \bibinfo{author}{\bibfnamefont{R.}~\bibnamefont{Jahn}},
  \bibinfo{journal}{Proceedings of the National Academy of Sciences}
  \textbf{\bibinfo{volume}{95}}, \bibinfo{pages}{15781} (\bibinfo{year}{1998}).

\bibitem[{\citenamefont{Heinrich and Rapoport}(2005)}]{heinrich2005generation}
\bibinfo{author}{\bibfnamefont{R.}~\bibnamefont{Heinrich}} \bibnamefont{and}
  \bibinfo{author}{\bibfnamefont{T.}~\bibnamefont{Rapoport}},
  \bibinfo{journal}{The Journal of Cell Biology}
  \textbf{\bibinfo{volume}{168}}, \bibinfo{pages}{271} (\bibinfo{year}{2005}).

\bibitem[{\citenamefont{Malsam and S{\"o}llner}(2011)}]{malsam2011organization}
\bibinfo{author}{\bibfnamefont{J.}~\bibnamefont{Malsam}} \bibnamefont{and}
  \bibinfo{author}{\bibfnamefont{T.}~\bibnamefont{S{\"o}llner}},
  \bibinfo{journal}{Cold Spring Harbor Perspectives in Biology}
  \textbf{\bibinfo{volume}{3}} (\bibinfo{year}{2011}).

\bibitem[{\citenamefont{Banfield et~al.}(1995)\citenamefont{Banfield, Lewis,
  and Pelham}}]{banfield1995snare}
\bibinfo{author}{\bibfnamefont{D.}~\bibnamefont{Banfield}},
  \bibinfo{author}{\bibfnamefont{M.}~\bibnamefont{Lewis}}, \bibnamefont{and}
  \bibinfo{author}{\bibfnamefont{H.}~\bibnamefont{Pelham}},
  \bibinfo{journal}{Nature} \textbf{\bibinfo{volume}{375}},
  \bibinfo{pages}{806} (\bibinfo{year}{1995}).

\bibitem[{\citenamefont{Lupashin and Sztul}(2005)}]{lupashin2005golgi}
\bibinfo{author}{\bibfnamefont{V.}~\bibnamefont{Lupashin}} \bibnamefont{and}
  \bibinfo{author}{\bibfnamefont{E.}~\bibnamefont{Sztul}},
  \bibinfo{journal}{Biochimica et Biophysica Acta (BBA)-Molecular Cell
  Research} \textbf{\bibinfo{volume}{1744}}, \bibinfo{pages}{325}
  (\bibinfo{year}{2005}).

\bibitem[{\citenamefont{Guo et~al.}(2008)\citenamefont{Guo, Punj, Sengupta, and
  Linstedt}}]{guo2008coat}
\bibinfo{author}{\bibfnamefont{Y.}~\bibnamefont{Guo}},
  \bibinfo{author}{\bibfnamefont{V.}~\bibnamefont{Punj}},
  \bibinfo{author}{\bibfnamefont{D.}~\bibnamefont{Sengupta}}, \bibnamefont{and}
  \bibinfo{author}{\bibfnamefont{A.}~\bibnamefont{Linstedt}},
  \bibinfo{journal}{Molecular Biology of the Cell}
  \textbf{\bibinfo{volume}{19}}, \bibinfo{pages}{2830} (\bibinfo{year}{2008}).

\bibitem[{\citenamefont{Ladinsky et~al.}(1999)\citenamefont{Ladinsky,
  Mastronarde, McIntosh, Howell, and Staehelin}}]{ladinsky1999golgi}
\bibinfo{author}{\bibfnamefont{M.}~\bibnamefont{Ladinsky}},
  \bibinfo{author}{\bibfnamefont{D.}~\bibnamefont{Mastronarde}},
  \bibinfo{author}{\bibfnamefont{J.}~\bibnamefont{McIntosh}},
  \bibinfo{author}{\bibfnamefont{K.}~\bibnamefont{Howell}}, \bibnamefont{and}
  \bibinfo{author}{\bibfnamefont{L.}~\bibnamefont{Staehelin}},
  \bibinfo{journal}{The Journal of Cell Biology}
  \textbf{\bibinfo{volume}{144}}, \bibinfo{pages}{1135} (\bibinfo{year}{1999}).

\bibitem[{\citenamefont{Weiss and Nilsson}(2000)}]{weiss2000protein}
\bibinfo{author}{\bibfnamefont{M.}~\bibnamefont{Weiss}} \bibnamefont{and}
  \bibinfo{author}{\bibfnamefont{T.}~\bibnamefont{Nilsson}},
  \bibinfo{journal}{FEBS Letters} \textbf{\bibinfo{volume}{486}},
  \bibinfo{pages}{2} (\bibinfo{year}{2000}).

\bibitem[{\citenamefont{Cosson et~al.}(2005)\citenamefont{Cosson, Ravazzola,
  Varlamov, S{\"o}llner, Di~Liberto, Volchuk, Rothman, and
  Orci}}]{cosson2005dynamic}
\bibinfo{author}{\bibfnamefont{P.}~\bibnamefont{Cosson}},
  \bibinfo{author}{\bibfnamefont{M.}~\bibnamefont{Ravazzola}},
  \bibinfo{author}{\bibfnamefont{O.}~\bibnamefont{Varlamov}},
  \bibinfo{author}{\bibfnamefont{T.}~\bibnamefont{S{\"o}llner}},
  \bibinfo{author}{\bibfnamefont{M.}~\bibnamefont{Di~Liberto}},
  \bibinfo{author}{\bibfnamefont{A.}~\bibnamefont{Volchuk}},
  \bibinfo{author}{\bibfnamefont{J.}~\bibnamefont{Rothman}}, \bibnamefont{and}
  \bibinfo{author}{\bibfnamefont{L.}~\bibnamefont{Orci}},
  \bibinfo{journal}{Proceedings of the National Academy of Sciences of the
  United States of America} \textbf{\bibinfo{volume}{102}},
  \bibinfo{pages}{14647} (\bibinfo{year}{2005}).

\bibitem[{\citenamefont{Wooding and Pelham}(1998)}]{wooding1998dynamics}
\bibinfo{author}{\bibfnamefont{S.}~\bibnamefont{Wooding}} \bibnamefont{and}
  \bibinfo{author}{\bibfnamefont{H.}~\bibnamefont{Pelham}},
  \bibinfo{journal}{Molecular Biology of the Cell}
  \textbf{\bibinfo{volume}{9}}, \bibinfo{pages}{2667} (\bibinfo{year}{1998}).

\bibitem[{\citenamefont{Xu et~al.}(2002)\citenamefont{Xu, Martin, James, and
  Hong}}]{xu2002gs15}
\bibinfo{author}{\bibfnamefont{Y.}~\bibnamefont{Xu}},
  \bibinfo{author}{\bibfnamefont{S.}~\bibnamefont{Martin}},
  \bibinfo{author}{\bibfnamefont{D.}~\bibnamefont{James}}, \bibnamefont{and}
  \bibinfo{author}{\bibfnamefont{W.}~\bibnamefont{Hong}},
  \bibinfo{journal}{Molecular Biology of the Cell}
  \textbf{\bibinfo{volume}{13}}, \bibinfo{pages}{3493} (\bibinfo{year}{2002}).

\bibitem[{\citenamefont{Bruinsma et~al.}(2004)\citenamefont{Bruinsma,
  Spelbrink, and Nothwehr}}]{bruinsma2004retrograde}
\bibinfo{author}{\bibfnamefont{P.}~\bibnamefont{Bruinsma}},
  \bibinfo{author}{\bibfnamefont{R.}~\bibnamefont{Spelbrink}},
  \bibnamefont{and} \bibinfo{author}{\bibfnamefont{S.}~\bibnamefont{Nothwehr}},
  \bibinfo{journal}{Journal of Biological Chemistry}
  \textbf{\bibinfo{volume}{279}}, \bibinfo{pages}{39814}
  (\bibinfo{year}{2004}).

\bibitem[{\citenamefont{Subramaniam et~al.}(1995)\citenamefont{Subramaniam,
  Krijnse-Locker, Tang, Ericsson, Yusoff, Griffiths, and
  Hong}}]{subramaniam1995monoclonal}
\bibinfo{author}{\bibfnamefont{V.}~\bibnamefont{Subramaniam}},
  \bibinfo{author}{\bibfnamefont{J.}~\bibnamefont{Krijnse-Locker}},
  \bibinfo{author}{\bibfnamefont{B.}~\bibnamefont{Tang}},
  \bibinfo{author}{\bibfnamefont{M.}~\bibnamefont{Ericsson}},
  \bibinfo{author}{\bibfnamefont{A.}~\bibnamefont{Yusoff}},
  \bibinfo{author}{\bibfnamefont{G.}~\bibnamefont{Griffiths}},
  \bibnamefont{and} \bibinfo{author}{\bibfnamefont{W.}~\bibnamefont{Hong}},
  \bibinfo{journal}{Journal of Cell Science} \textbf{\bibinfo{volume}{108}},
  \bibinfo{pages}{2405} (\bibinfo{year}{1995}).

\bibitem[{\citenamefont{Malsam et~al.}(2008)\citenamefont{Malsam, Kreye, and
  Sollner}}]{malsam2008membrane}
\bibinfo{author}{\bibfnamefont{J.}~\bibnamefont{Malsam}},
  \bibinfo{author}{\bibfnamefont{S.}~\bibnamefont{Kreye}}, \bibnamefont{and}
  \bibinfo{author}{\bibfnamefont{T.}~\bibnamefont{Sollner}},
  \bibinfo{journal}{Cellular and Molecular Life Sciences CMLS}
  \textbf{\bibinfo{volume}{65}}, \bibinfo{pages}{2814} (\bibinfo{year}{2008}).

\bibitem[{\citenamefont{Miller et~al.}(2003)\citenamefont{Miller, Beilharz,
  Malkus, Lee, Hamamoto, Orci, and Schekman}}]{miller2003multiple}
\bibinfo{author}{\bibfnamefont{E.~A.} \bibnamefont{Miller}},
  \bibinfo{author}{\bibfnamefont{T.~H.} \bibnamefont{Beilharz}},
  \bibinfo{author}{\bibfnamefont{P.~N.} \bibnamefont{Malkus}},
  \bibinfo{author}{\bibfnamefont{M.}~\bibnamefont{Lee}},
  \bibinfo{author}{\bibfnamefont{S.}~\bibnamefont{Hamamoto}},
  \bibinfo{author}{\bibfnamefont{L.}~\bibnamefont{Orci}}, \bibnamefont{and}
  \bibinfo{author}{\bibfnamefont{R.}~\bibnamefont{Schekman}},
  \bibinfo{journal}{Cell} \textbf{\bibinfo{volume}{114}}, \bibinfo{pages}{497}
  (\bibinfo{year}{2003}).

\bibitem[{\citenamefont{Jarvela and Linstedt}(2012)}]{jarvela2012irradiation}
\bibinfo{author}{\bibfnamefont{T.}~\bibnamefont{Jarvela}} \bibnamefont{and}
  \bibinfo{author}{\bibfnamefont{A.}~\bibnamefont{Linstedt}},
  \bibinfo{journal}{Journal of Cell Science} \textbf{\bibinfo{volume}{125}},
  \bibinfo{pages}{973} (\bibinfo{year}{2012}).

\bibitem[{\citenamefont{Zhang and Hong}(2001)}]{zhang2001ykt6}
\bibinfo{author}{\bibfnamefont{T.}~\bibnamefont{Zhang}} \bibnamefont{and}
  \bibinfo{author}{\bibfnamefont{W.}~\bibnamefont{Hong}},
  \bibinfo{journal}{Journal of Biological Chemistry}
  \textbf{\bibinfo{volume}{276}}, \bibinfo{pages}{27480}
  (\bibinfo{year}{2001}).

\bibitem[{\citenamefont{Hui et~al.}(1997)\citenamefont{Hui, Nakamura,
  S{\"o}nnichsen, Shima, Nilsson, and Warren}}]{hui1997isoform}
\bibinfo{author}{\bibfnamefont{N.}~\bibnamefont{Hui}},
  \bibinfo{author}{\bibfnamefont{N.}~\bibnamefont{Nakamura}},
  \bibinfo{author}{\bibfnamefont{B.}~\bibnamefont{S{\"o}nnichsen}},
  \bibinfo{author}{\bibfnamefont{D.}~\bibnamefont{Shima}},
  \bibinfo{author}{\bibfnamefont{T.}~\bibnamefont{Nilsson}}, \bibnamefont{and}
  \bibinfo{author}{\bibfnamefont{G.}~\bibnamefont{Warren}},
  \bibinfo{journal}{Molecular Biology of the Cell}
  \textbf{\bibinfo{volume}{8}}, \bibinfo{pages}{1777} (\bibinfo{year}{1997}).

\bibitem[{\citenamefont{Preuss et~al.}(1992)\citenamefont{Preuss, Mulholland,
  Franzusoff, Segev, and Botstein}}]{preuss1992characterization}
\bibinfo{author}{\bibfnamefont{D.}~\bibnamefont{Preuss}},
  \bibinfo{author}{\bibfnamefont{J.}~\bibnamefont{Mulholland}},
  \bibinfo{author}{\bibfnamefont{A.}~\bibnamefont{Franzusoff}},
  \bibinfo{author}{\bibfnamefont{N.}~\bibnamefont{Segev}}, \bibnamefont{and}
  \bibinfo{author}{\bibfnamefont{D.}~\bibnamefont{Botstein}},
  \bibinfo{journal}{Molecular Biology of the Cell}
  \textbf{\bibinfo{volume}{3}}, \bibinfo{pages}{789} (\bibinfo{year}{1992}).

\bibitem[{\citenamefont{Losev et~al.}(2006)\citenamefont{Losev, Reinke, Jellen,
  Strongin, Bevis, and Glick}}]{losev2006golgi}
\bibinfo{author}{\bibfnamefont{E.}~\bibnamefont{Losev}},
  \bibinfo{author}{\bibfnamefont{C.}~\bibnamefont{Reinke}},
  \bibinfo{author}{\bibfnamefont{J.}~\bibnamefont{Jellen}},
  \bibinfo{author}{\bibfnamefont{D.}~\bibnamefont{Strongin}},
  \bibinfo{author}{\bibfnamefont{B.}~\bibnamefont{Bevis}}, \bibnamefont{and}
  \bibinfo{author}{\bibfnamefont{B.}~\bibnamefont{Glick}},
  \bibinfo{journal}{Nature} \textbf{\bibinfo{volume}{441}},
  \bibinfo{pages}{1002} (\bibinfo{year}{2006}).

\bibitem[{\citenamefont{Matsuura-Tokita
  et~al.}(2006)\citenamefont{Matsuura-Tokita, Takeuchi, Ichihara, Mikuriya, and
  Nakano}}]{matsuura2006live}
\bibinfo{author}{\bibfnamefont{K.}~\bibnamefont{Matsuura-Tokita}},
  \bibinfo{author}{\bibfnamefont{M.}~\bibnamefont{Takeuchi}},
  \bibinfo{author}{\bibfnamefont{A.}~\bibnamefont{Ichihara}},
  \bibinfo{author}{\bibfnamefont{K.}~\bibnamefont{Mikuriya}}, \bibnamefont{and}
  \bibinfo{author}{\bibfnamefont{A.}~\bibnamefont{Nakano}},
  \bibinfo{journal}{Nature} \textbf{\bibinfo{volume}{441}},
  \bibinfo{pages}{1007} (\bibinfo{year}{2006}).

\bibitem[{\citenamefont{Rivera-Molina and Novick}(2009)}]{rivera2009rab}
\bibinfo{author}{\bibfnamefont{F.}~\bibnamefont{Rivera-Molina}}
  \bibnamefont{and} \bibinfo{author}{\bibfnamefont{P.}~\bibnamefont{Novick}},
  \bibinfo{journal}{Proceedings of the National Academy of Sciences}
  \textbf{\bibinfo{volume}{106}}, \bibinfo{pages}{14408}
  (\bibinfo{year}{2009}).

\bibitem[{\citenamefont{Papanikou and Glick}(2009)}]{papanikou2009yeast}
\bibinfo{author}{\bibfnamefont{E.}~\bibnamefont{Papanikou}} \bibnamefont{and}
  \bibinfo{author}{\bibfnamefont{B.}~\bibnamefont{Glick}},
  \bibinfo{journal}{FEBS Letters} \textbf{\bibinfo{volume}{583}},
  \bibinfo{pages}{3746} (\bibinfo{year}{2009}).

\bibitem[{\citenamefont{Okamoto et~al.}(2012)\citenamefont{Okamoto, Kurokawa,
  Matsuura-Tokita, Saito, Hirata, and Nakano}}]{okamoto2012high}
\bibinfo{author}{\bibfnamefont{M.}~\bibnamefont{Okamoto}},
  \bibinfo{author}{\bibfnamefont{K.}~\bibnamefont{Kurokawa}},
  \bibinfo{author}{\bibfnamefont{K.}~\bibnamefont{Matsuura-Tokita}},
  \bibinfo{author}{\bibfnamefont{C.}~\bibnamefont{Saito}},
  \bibinfo{author}{\bibfnamefont{R.}~\bibnamefont{Hirata}}, \bibnamefont{and}
  \bibinfo{author}{\bibfnamefont{A.}~\bibnamefont{Nakano}},
  \bibinfo{journal}{Journal of Cell Science} \textbf{\bibinfo{volume}{125}},
  \bibinfo{pages}{3412} (\bibinfo{year}{2012}).

\bibitem[{\citenamefont{Sagiv et~al.}(2000)\citenamefont{Sagiv, Legesse-Miller,
  Porat, and Elazar}}]{sagiv2000gate}
\bibinfo{author}{\bibfnamefont{Y.}~\bibnamefont{Sagiv}},
  \bibinfo{author}{\bibfnamefont{A.}~\bibnamefont{Legesse-Miller}},
  \bibinfo{author}{\bibfnamefont{A.}~\bibnamefont{Porat}}, \bibnamefont{and}
  \bibinfo{author}{\bibfnamefont{Z.}~\bibnamefont{Elazar}},
  \bibinfo{journal}{The EMBO Journal} \textbf{\bibinfo{volume}{19}},
  \bibinfo{pages}{1494} (\bibinfo{year}{2000}).

\bibitem[{\citenamefont{Zhong et~al.}(2011)\citenamefont{Zhong, Zhou, Li, Zhou,
  Ma, Wei, Li, Olkkonen, and Yan}}]{zhong2011osbp}
\bibinfo{author}{\bibfnamefont{W.}~\bibnamefont{Zhong}},
  \bibinfo{author}{\bibfnamefont{Y.}~\bibnamefont{Zhou}},
  \bibinfo{author}{\bibfnamefont{S.}~\bibnamefont{Li}},
  \bibinfo{author}{\bibfnamefont{T.}~\bibnamefont{Zhou}},
  \bibinfo{author}{\bibfnamefont{H.}~\bibnamefont{Ma}},
  \bibinfo{author}{\bibfnamefont{K.}~\bibnamefont{Wei}},
  \bibinfo{author}{\bibfnamefont{H.}~\bibnamefont{Li}},
  \bibinfo{author}{\bibfnamefont{V.}~\bibnamefont{Olkkonen}}, \bibnamefont{and}
  \bibinfo{author}{\bibfnamefont{D.}~\bibnamefont{Yan}},
  \bibinfo{journal}{Experimental Cell Research} \textbf{\bibinfo{volume}{317}},
  \bibinfo{pages}{2353} (\bibinfo{year}{2011}).

\bibitem[{\citenamefont{Oka et~al.}(2004)\citenamefont{Oka, Ungar, Hughson, and
  Krieger}}]{oka2004cog}
\bibinfo{author}{\bibfnamefont{T.}~\bibnamefont{Oka}},
  \bibinfo{author}{\bibfnamefont{D.}~\bibnamefont{Ungar}},
  \bibinfo{author}{\bibfnamefont{F.}~\bibnamefont{Hughson}}, \bibnamefont{and}
  \bibinfo{author}{\bibfnamefont{M.}~\bibnamefont{Krieger}},
  \bibinfo{journal}{Molecular Biology of the Cell}
  \textbf{\bibinfo{volume}{15}}, \bibinfo{pages}{2423} (\bibinfo{year}{2004}).

\bibitem[{\citenamefont{Ohashi and Munro}(2010)}]{ohashi2010membrane}
\bibinfo{author}{\bibfnamefont{Y.}~\bibnamefont{Ohashi}} \bibnamefont{and}
  \bibinfo{author}{\bibfnamefont{S.}~\bibnamefont{Munro}},
  \bibinfo{journal}{Molecular Biology of the Cell}
  \textbf{\bibinfo{volume}{21}}, \bibinfo{pages}{3998} (\bibinfo{year}{2010}).

\bibitem[{\citenamefont{Yamamoto et~al.}(2012)\citenamefont{Yamamoto, Kakuta,
  Watanabe, Kitamura, Sekito, Kondo-Kakuta, Ichikawa, Kinjo, and
  Ohsumi}}]{yamamoto2012atg9}
\bibinfo{author}{\bibfnamefont{H.}~\bibnamefont{Yamamoto}},
  \bibinfo{author}{\bibfnamefont{S.}~\bibnamefont{Kakuta}},
  \bibinfo{author}{\bibfnamefont{T.}~\bibnamefont{Watanabe}},
  \bibinfo{author}{\bibfnamefont{A.}~\bibnamefont{Kitamura}},
  \bibinfo{author}{\bibfnamefont{T.}~\bibnamefont{Sekito}},
  \bibinfo{author}{\bibfnamefont{C.}~\bibnamefont{Kondo-Kakuta}},
  \bibinfo{author}{\bibfnamefont{R.}~\bibnamefont{Ichikawa}},
  \bibinfo{author}{\bibfnamefont{M.}~\bibnamefont{Kinjo}}, \bibnamefont{and}
  \bibinfo{author}{\bibfnamefont{Y.}~\bibnamefont{Ohsumi}},
  \bibinfo{journal}{The Journal of Cell Biology}
  \textbf{\bibinfo{volume}{198}}, \bibinfo{pages}{219} (\bibinfo{year}{2012}).

\bibitem[{\citenamefont{Lang et~al.}(2002)\citenamefont{Lang, Margittai,
  H{\"o}lzler, and Jahn}}]{lang2002snares}
\bibinfo{author}{\bibfnamefont{T.}~\bibnamefont{Lang}},
  \bibinfo{author}{\bibfnamefont{M.}~\bibnamefont{Margittai}},
  \bibinfo{author}{\bibfnamefont{H.}~\bibnamefont{H{\"o}lzler}},
  \bibnamefont{and} \bibinfo{author}{\bibfnamefont{R.}~\bibnamefont{Jahn}},
  \bibinfo{journal}{The Journal of Cell Biology}
  \textbf{\bibinfo{volume}{158}}, \bibinfo{pages}{751} (\bibinfo{year}{2002}).

\bibitem[{\citenamefont{Bar-On et~al.}(2008)\citenamefont{Bar-On, Winter,
  Nachliel, Gutman, Fasshauer, Lang, and Ashery}}]{bar2008imaging}
\bibinfo{author}{\bibfnamefont{D.}~\bibnamefont{Bar-On}},
  \bibinfo{author}{\bibfnamefont{U.}~\bibnamefont{Winter}},
  \bibinfo{author}{\bibfnamefont{E.}~\bibnamefont{Nachliel}},
  \bibinfo{author}{\bibfnamefont{M.}~\bibnamefont{Gutman}},
  \bibinfo{author}{\bibfnamefont{D.}~\bibnamefont{Fasshauer}},
  \bibinfo{author}{\bibfnamefont{T.}~\bibnamefont{Lang}}, \bibnamefont{and}
  \bibinfo{author}{\bibfnamefont{U.}~\bibnamefont{Ashery}},
  \bibinfo{journal}{FEBS Letters} \textbf{\bibinfo{volume}{582}},
  \bibinfo{pages}{3563} (\bibinfo{year}{2008}).

\bibitem[{\citenamefont{Storrie et~al.}(1998)\citenamefont{Storrie, White,
  R{\"o}ttger, Stelzer, Suganuma, and Nilsson}}]{storrie1998recycling}
\bibinfo{author}{\bibfnamefont{B.}~\bibnamefont{Storrie}},
  \bibinfo{author}{\bibfnamefont{J.}~\bibnamefont{White}},
  \bibinfo{author}{\bibfnamefont{S.}~\bibnamefont{R{\"o}ttger}},
  \bibinfo{author}{\bibfnamefont{E.}~\bibnamefont{Stelzer}},
  \bibinfo{author}{\bibfnamefont{T.}~\bibnamefont{Suganuma}}, \bibnamefont{and}
  \bibinfo{author}{\bibfnamefont{T.}~\bibnamefont{Nilsson}},
  \bibinfo{journal}{The Journal of Cell Biology}
  \textbf{\bibinfo{volume}{143}}, \bibinfo{pages}{1505} (\bibinfo{year}{1998}).

\bibitem[{\citenamefont{Lewis and Pelham}(1996)}]{lewis1996snare}
\bibinfo{author}{\bibfnamefont{M.~J.} \bibnamefont{Lewis}} \bibnamefont{and}
  \bibinfo{author}{\bibfnamefont{H.~R.} \bibnamefont{Pelham}},
  \bibinfo{journal}{Cell} \textbf{\bibinfo{volume}{85}}, \bibinfo{pages}{205}
  (\bibinfo{year}{1996}).

\bibitem[{\citenamefont{Lewis et~al.}(1997)\citenamefont{Lewis, Rayner, and
  Pelham}}]{lewis1997novel}
\bibinfo{author}{\bibfnamefont{M.}~\bibnamefont{Lewis}},
  \bibinfo{author}{\bibfnamefont{J.}~\bibnamefont{Rayner}}, \bibnamefont{and}
  \bibinfo{author}{\bibfnamefont{H.}~\bibnamefont{Pelham}},
  \bibinfo{journal}{The EMBO Journal} \textbf{\bibinfo{volume}{16}},
  \bibinfo{pages}{3017} (\bibinfo{year}{1997}).

\bibitem[{\citenamefont{Kumar and Gromiha}(2006)}]{kumar2006pint}
\bibinfo{author}{\bibfnamefont{M.~S.} \bibnamefont{Kumar}} \bibnamefont{and}
  \bibinfo{author}{\bibfnamefont{M.~M.} \bibnamefont{Gromiha}},
  \bibinfo{journal}{Nucleic Acids Research} \textbf{\bibinfo{volume}{34}},
  \bibinfo{pages}{D195} (\bibinfo{year}{2006}).

\bibitem[{\citenamefont{Gannon et~al.}(2011)\citenamefont{Gannon, Bergeron, and
  Nilsson}}]{gannon2011golgi}
\bibinfo{author}{\bibfnamefont{J.}~\bibnamefont{Gannon}},
  \bibinfo{author}{\bibfnamefont{J.}~\bibnamefont{Bergeron}}, \bibnamefont{and}
  \bibinfo{author}{\bibfnamefont{T.}~\bibnamefont{Nilsson}},
  \bibinfo{journal}{Cold Spring Harbor Perspectives in Biology}
  \textbf{\bibinfo{volume}{3}} (\bibinfo{year}{2011}).

\bibitem[{\citenamefont{Tai et~al.}(2004)\citenamefont{Tai, Lu, Wang, Tang,
  Goud, Johannes, and Hong}}]{tai2004participation}
\bibinfo{author}{\bibfnamefont{G.}~\bibnamefont{Tai}},
  \bibinfo{author}{\bibfnamefont{L.}~\bibnamefont{Lu}},
  \bibinfo{author}{\bibfnamefont{T.}~\bibnamefont{Wang}},
  \bibinfo{author}{\bibfnamefont{B.}~\bibnamefont{Tang}},
  \bibinfo{author}{\bibfnamefont{B.}~\bibnamefont{Goud}},
  \bibinfo{author}{\bibfnamefont{L.}~\bibnamefont{Johannes}}, \bibnamefont{and}
  \bibinfo{author}{\bibfnamefont{W.}~\bibnamefont{Hong}},
  \bibinfo{journal}{Molecular Biology of the Cell}
  \textbf{\bibinfo{volume}{15}}, \bibinfo{pages}{4011} (\bibinfo{year}{2004}).

\bibitem[{\citenamefont{Staehelin and Kang}(2008)}]{staehelin2008nanoscale}
\bibinfo{author}{\bibfnamefont{L.}~\bibnamefont{Staehelin}} \bibnamefont{and}
  \bibinfo{author}{\bibfnamefont{B.}~\bibnamefont{Kang}},
  \bibinfo{journal}{Plant Physiology} \textbf{\bibinfo{volume}{147}},
  \bibinfo{pages}{1454} (\bibinfo{year}{2008}).

\bibitem[{\citenamefont{Lanoix et~al.}(2001)\citenamefont{Lanoix, Ouwendijk,
  Stark, Szafer, Cassel, Dejgaard, Weiss, and Nilsson}}]{lanoix2001sorting}
\bibinfo{author}{\bibfnamefont{J.}~\bibnamefont{Lanoix}},
  \bibinfo{author}{\bibfnamefont{J.}~\bibnamefont{Ouwendijk}},
  \bibinfo{author}{\bibfnamefont{A.}~\bibnamefont{Stark}},
  \bibinfo{author}{\bibfnamefont{E.}~\bibnamefont{Szafer}},
  \bibinfo{author}{\bibfnamefont{D.}~\bibnamefont{Cassel}},
  \bibinfo{author}{\bibfnamefont{K.}~\bibnamefont{Dejgaard}},
  \bibinfo{author}{\bibfnamefont{M.}~\bibnamefont{Weiss}}, \bibnamefont{and}
  \bibinfo{author}{\bibfnamefont{T.}~\bibnamefont{Nilsson}},
  \bibinfo{journal}{The Journal of Cell Biology}
  \textbf{\bibinfo{volume}{155}}, \bibinfo{pages}{1199} (\bibinfo{year}{2001}).

\bibitem[{\citenamefont{Mart{\'\i}nez-Alonso
  et~al.}(2007)\citenamefont{Mart{\'\i}nez-Alonso, Ballesta, and
  Mart{\'\i}nez-Men{\'a}rguez}}]{martinez2007low}
\bibinfo{author}{\bibfnamefont{E.}~\bibnamefont{Mart{\'\i}nez-Alonso}},
  \bibinfo{author}{\bibfnamefont{J.}~\bibnamefont{Ballesta}}, \bibnamefont{and}
  \bibinfo{author}{\bibfnamefont{J.}~\bibnamefont{Mart{\'\i}nez-Men{\'a}rguez}},
  \bibinfo{journal}{Traffic} \textbf{\bibinfo{volume}{8}}, \bibinfo{pages}{359}
  (\bibinfo{year}{2007}).

\bibitem[{\citenamefont{Moelleken et~al.}(2007)\citenamefont{Moelleken, Malsam,
  Betts, Movafeghi, Reckmann, Meissner, Hellwig, Russell, S{\"o}llner,
  Br{\"u}gger et~al.}}]{moelleken2007differential}
\bibinfo{author}{\bibfnamefont{J.}~\bibnamefont{Moelleken}},
  \bibinfo{author}{\bibfnamefont{J.}~\bibnamefont{Malsam}},
  \bibinfo{author}{\bibfnamefont{M.}~\bibnamefont{Betts}},
  \bibinfo{author}{\bibfnamefont{A.}~\bibnamefont{Movafeghi}},
  \bibinfo{author}{\bibfnamefont{I.}~\bibnamefont{Reckmann}},
  \bibinfo{author}{\bibfnamefont{I.}~\bibnamefont{Meissner}},
  \bibinfo{author}{\bibfnamefont{A.}~\bibnamefont{Hellwig}},
  \bibinfo{author}{\bibfnamefont{R.}~\bibnamefont{Russell}},
  \bibinfo{author}{\bibfnamefont{T.}~\bibnamefont{S{\"o}llner}},
  \bibinfo{author}{\bibfnamefont{B.}~\bibnamefont{Br{\"u}gger}},
  \bibnamefont{et~al.}, \bibinfo{journal}{Proceedings of the National Academy
  of Sciences} \textbf{\bibinfo{volume}{104}}, \bibinfo{pages}{4425}
  (\bibinfo{year}{2007}).

\end{thebibliography}

\end{document}